\documentclass[a4paper,11pt]{article}
\pdfoutput=1 

\usepackage{jcappub} 

\usepackage[T1]{fontenc} 
\usepackage{placeins}
\usepackage[dvipsnames,svgnames,x11names,hyperref]{xcolor}
\usepackage{array}
\usepackage{graphicx}
\usepackage{booktabs}
\usepackage{slashed}
\RequirePackage{mathrsfs}
\usepackage{dsfont}
\usepackage{hyperref}
\hypersetup{
        unicode = true,
        pdftitle = Note , 
        colorlinks = true,
        linkcolor = Black!30!DodgerBlue4,
        citecolor = Black!30!DodgerBlue4,
        filecolor = Black!30!DodgerBlue4,
        urlcolor = Black!30!DodgerBlue4,
}

\usepackage{booktabs}

\graphicspath{{./plots/}}

\usepackage{floatrow}
\newfloatcommand{capbtabbox}{table}[][\FBwidth]

\newcommand{\erf}{\mathop{\mathrm{erf}}}

\newcommand{\dd}{\mathrm{d}}

\renewcommand{\epsilon}{\varepsilon}

\renewcommand{\bar}{\overline}

\newcolumntype{L}{>{$}l<{$}}
\newcolumntype{C}{>{$}c<{$}}

\newcommand{\vPerpEl}{\mathbf{v}_{\rm el}^\perp}
\allowdisplaybreaks

\title{\boldmath Rejecting the Majorana nature of dark matter with electron scattering experiments}

\author[a]{Riccardo Catena}
\author[a]{Timon Emken}
\author[a]{and Julia Ravanis}

\affiliation[a]{Chalmers University of Technology, Department of Physics, SE-412 96 G\"oteborg, Sweden}

\emailAdd{catena@chalmers.se}
\emailAdd{emken@chalmers.se}
\emailAdd{julia.ravanis@chalmers.se}

\abstract{Assuming that Dark Matter (DM) is made of fermions in the sub-GeV mass range with interactions dominated by electromagnetic moments of higher order, such as the electric and magnetic dipoles or the anapole moment, we show that direct detection experiments searching for atomic ionisation events in xenon targets can shed light on whether DM is a Dirac or Majorana particle.~Specifically, we find that between about 45 (120) and 610 (1700) signal events are required to reject Majorana DM in favour of Dirac DM with a statistical significance corresponding to 3 (5) standard deviations.~The exact number of DM signal events corresponding to a given significance depends on the relative size of the anapole, magnetic dipole and electric dipole contributions to the expected rate of DM-induced atomic ionisations under the Dirac hypothesis.~Our conclusions are based on Monte Carlo simulations and the likelihood ratio test.~While the use of asymptotic formulae for the latter is standard in many applications, here it requires a non-trivial extension to the case where one of the hypotheses lies on the boundary of the parameter space.~Our results constitute a solid proof of concept about the possibility of using direct detection experiments to reject the Majorana DM hypothesis when the DM interactions are dominated by higher-order electromagnetic moments.
} 

\begin{document} 
\maketitle
\flushbottom

\section{Introduction}
Despite impressive experimental efforts, the particles forming our Universe's Dark Matter (DM) component have continued escaping detection for about four decades~\cite{Bertone:2018xtm}.~This has motivated a critical reassessment of the standard DM paradigm, leading to the rise of a new framework where DM is made of particles lighter than the lightest nuclei and communicates with the visible world via interactions with electrons, rather than with nuclei~\cite{Alexander:2016aln}.~The search for light, or sub-GeV, DM candidates mainly relies on the observation of electronic transitions in condensed matter systems, including dual-phase argon~\cite{Agnes:2018oej} and xenon~\cite{Essig:2012yx,Essig:2017kqs,Aprile:2019xxb} targets, superconductors \cite{Hochberg:2015pha,Hochberg:2015fth,Hochberg:2019cyy}, graphene~\cite{Hochberg:2016ntt,Geilhufe:2018gry}, 3D Dirac materials~\cite{Hochberg:2017wce,Geilhufe:2019ndy,Coskuner:2019odd}, scintillators~\cite{Derenzo:2016fse,Blanco:2019lrf} and polar crystals~\cite{Knapen:2017ekk}.~In this mass range, electronic transitions are kinematically favoured compared to nuclear recoils~\cite{Petricca:2017zdp,Aprile:2019jmx}.~Theoretical efforts in modelling sub-GeV DM are reviewed in~\cite{Battaglieri:2017aum}, while the performances of different materials in the search for sub-GeV DM are compared in~\cite{Griffin:2019mvc}.

When testing the predictions of models for sub-GeV DM against the current null result of DM direct detection experiments, e.g.~\cite{Agnese:2018col,Crisler:2018gci,Aguilar-Arevalo:2019wdi,Abramoff:2019dfb}, or computing the projected sensitivity of next-generation DM search experiments~\cite{Essig:2015cda,Andersson:2020uwc}, knowing how condensed matter systems ``respond'' to DM-electron interactions is crucial.~Specifically, the rate of DM-induced electronic transitions in condensed matter systems is proportional to a linear combination of electron wave function overlap integrals known as ``response functions''~\cite{Catena:2019gfa}.~Recently, it has been pointed out that argon and xenon targets can respond in four different ways when ionised by using DM as an external probe~\cite{Catena:2019gfa}.~One of these response functions is the standard ``ionisation form factor'', familiar from the literature on sub-GeV DM, e.g.~\cite{Kopp:2009et,Essig:2011nj}.~The remaining three atomic response functions were identified in~\cite{Catena:2019gfa} for the first time and describe distortions in the ionisation spectrum induced by the finite dispersion of the initial state electron momentum distribution.

The results found in~\cite{Catena:2019gfa} have multiple implications.~On the one hand, they show that there are material properties that have so far remained ``hidden'' and that can only be revealed if DM is used as an external probe.~On the other hand, they allow us to interpret the results reported by the XENON~\cite{Aprile:2019xxb} and Darkside~\cite{Agnes:2018oej} collaborations within models for sub-GeV DM that, so far, were only tractable within simplifying approximations~\cite{Graham:2012su,Essig:2012yx,Chang:2019xva}.~Models where DM is electrically neutral but still interacts with the Standard Model via higher-order electromagnetic moments provide a prominent example of scenarios that cannot be investigated without the atomic response functions found in~\cite{Catena:2019gfa}.~Within this framework, DM can interact with electrons via the electric dipole, the magnetic dipole and the electromagnetic anapole coupling (see~\cite{Kavanagh:2018xeh} for a general classification of DM-photon couplings and higher-order electromagnetic moments).~In general, the novel response functions found in~\cite{Catena:2019gfa} arise whenever the amplitude for DM-electron scattering explicitly depends on the momentum of the initial state electron.

The higher-order electromagnetic moments of spin-1/2 DM are extremely interesting as their amplitude depends on whether DM is a Majorana or a Dirac particle.~The leading electromagnetic moment for Majorana DM is the anapole moment.~Electric and magnetic dipoles vanish exactly in this case.~For Dirac DM, instead, anapole, magnetic dipole and electric dipole moments can be simultaneously different from zero.~This observation allows us to formulate a ``Dirac DM hypothesis'', where DM is a particle with anapole, electric and magnetic dipole interactions (and, for simplicity, no other interactions), and a ``Majorana DM hypothesis'', where the DM candidate has anapole interactions only.~It is then natural to ask whether the observation of higher-order electromagnetic moments can be used to statistically discriminate one hypothesis from the other, and if so, with what significance.~A similar question was raised in Refs.~\cite{Queiroz:2016sxf,Kavanagh:2017hcl} focusing on scalar and vector couplings between DM and nuclei.

The aim of this work is to statistically compare the Dirac and Majorana DM hypotheses in the light of a future discovery of DM at direct detection experiments using xenon as a target material.~Since our focus is on sub-GeV DM, we assume that such discovery occurred via the observation of DM-induced atomic ionisations.~This analysis is motivated by recent theoretical advances~\cite{Catena:2019gfa} which, for the first time, enable us to model the anapole, electric dipole and magnetic dipole interactions in a rigorous manner.~We compare the two hypotheses by performing Monte Carlo simulations of atomic ionisation data from next-generation xenon experiments and using the likelihood ratio, and its asymptotic expansion in terms of a chi-bar-square distribution~($\bar{\chi}^2$), as a test statistic.~This is the correct procedure when, as in the present analysis, one of the tested hypotheses lies on the boundary of the parameter space~\cite{Shapiro1985Apr}.~We present our results in terms of number of signal events required to reject the ``null'', Majorana hypothesis in favour of the alternative, Dirac hypothesis at a given statistical significance.~Focusing on selected benchmark values for the DM coupling constants, we highlight in what regions of the parameter space the significance for rejecting the Majorana hypothesis is higher.

This paper is organised as follows.~In Sec.~\ref{sec:theory}, we introduce the leading electromagnetic moments of light DM, formulating a Majorana DM hypothesis and a Dirac DM hypothesis.~For each hypothesis, we provide interaction Lagrangian and amplitude for DM-electron scattering.~We then compute the corresponding rate of DM-induced atomic ionisations in Sec.~\ref{sec:ionisation}.~In Sec.~\ref{sec:stat}, we introduce the statistical framework that we use to compare the Dirac and Majorana DM hypotheses formulated in Sec.~\ref{sec:theory}.~We present the results of this comparison in Sec.~\ref{sec:results} and conclude in Sec.~\ref{sec:conclusions}.~In the Appendix, we provide a detailed derivation of the amplitude for DM-electron scattering for the anapole, magnetic and electric dipole DM couplings. Finally, we provide the code used to perform the statistical analysis~\cite{Emken2020}, which is archived as~\href{https://doi.org/10.5281/zenodo.3701262}{[DOI:10.5281/zenodo.3701262]}.

\section{Dark matter with higher-order electromagnetic moments}
\label{sec:theory}
We are interested in models for Majorana and Dirac DM where the DM particle couples to the Standard Model photon via higher-order electromagnetic moments:~the magnetic and electric dipole and the electromagnetic anapole~\cite{Kavanagh:2018xeh}.~This theoretical framework will allow us to formulate testable Dirac and Majorana DM hypotheses.~For each model introduced here, we provide interaction Lagrangian and amplitude for DM-electron scattering.~A derivation of these expressions can be found in App.~\ref{sec:amplitudes}.~We apply the results in App.~\ref{sec:amplitudes} to compute the rate of DM-induced atomic ionisations in Sec.~\ref{sec:ionisation}.~In Sec.~\ref{sec:stat}, we use this rate as a physical observable to test the Dirac DM hypothesis against the Majorana DM hypothesis with atomic ionisation data.

\subsection{Majorana dark matter}
In the case of Majorana DM, magnetic and electric dipole interactions are identically zero, as these moments are odd under particle-antiparticle exchange.~The anapole operator, however, does not vanish because it is even under the same transformation.~The associated interaction Lagrangian, $\mathscr{L}_I^{(M)}$, is therefore given by the anapole term only,
\begin{align}
\mathscr{L}_I^{(M)} = \mathscr{L}_a^{(M)} \equiv \frac{1}{2}\frac{g_1}{\Lambda^2} \, \bar{\chi} \gamma^\mu \gamma^5 \chi \,\partial^{\nu} F_{\mu \nu} \,,
\label{eq:LIa}
\end{align}
where $\chi$ is a four-component spinor field for the Majorana DM particle, $F_{\mu\nu}$ is the electromagnetic field strength tensor, $g_1$ is a dimensionless coupling constant and $\Lambda$ is a mass scale.~Here, we assume that the electromagnetic moments of DM are generated at the scale they are measured at direct detection experiment (i.e.~below 1 GeV) and do not consider renormalisation group effects and the associated ultraviolet completion.~For a detailed discussion on these aspects, we refer to~\cite{Kavanagh:2018xeh}.~We use the Lagrangian in Eq.~(\ref{eq:LIa}) to compute the amplitude for DM-electron scattering and the associated rate of DM-induced atomic ionisations.~For Majorana DM with anapole interactions, the non-relativistic amplitude for DM-electron scattering can be expressed in terms of tridimensional momentum transfer, $\mathbf{q}$, transverse relative velocity $\mathbf{v}_{\rm el}^\perp$ (defined below in Sec.~\ref{sec:ionisation}), and the electron and DM particle spin, $\mathbf{S}_e$ and $\mathbf{S}_\chi$, respectively, 
\begin{align}
\mathcal{M} = \frac{4 e g_1}{\Lambda^2} m_\chi m_e \Bigg\{
    2 \left(\mathbf{v}_{\rm el}^\perp \cdot \xi^{ s'\dagger} \mathbf{S}_\chi \xi^s \right) \delta^{r' r} +
    g_e \left( \xi^{s'\dagger } \mathbf{S}_\chi \xi^s \right) \cdot \left( i\frac{\mathbf{q}}{m_e} \times \eta^{r'\dagger} \mathbf{S}_e \eta^r  \right) 
    \Bigg\}\,, 
\label{eq:MNRa}
\end{align}
where $m_\chi$ and $m_e$ are the DM particle and electron rest mass, $\mathbf{S}_e=\boldsymbol{\sigma}/2$ ($\mathbf{S}_\chi=\boldsymbol{\sigma}/2$) is the spin operator in the electron (DM) spin space and $\boldsymbol{\sigma}=(\sigma_1,\sigma_2,\sigma_3)$ are the Pauli matrices.~In Eq.~(\ref{eq:MNRa}), $\eta^r$, $r=1,2$ ($\xi^s$, $s=1,2$) are two-component spinors acting on the electron (DM) spin space.~The electron $g$-factor, $g_e\approx 2$, is defined in the Appendix in terms of electromagnetic form factors.

\subsection{Dirac dark matter}
In the case of Dirac DM, anapole, magnetic dipole and electric dipole moments can all be different from zero, as the underlying theory for DM does not have to be symmetric under particle-antiparticle exchange.~In principle, the three moments can simultaneously participate in the interaction between DM and electrons.~However, we will find that there is no interference between different moments.~For Dirac DM, the interaction Lagrangian associated with the anapole moment is 
\begin{align}
\mathscr{L}^{(D)}_a = \frac{g_1}{\Lambda^2} \, \bar{\psi} \gamma^\mu \gamma^5 \psi \,\partial^{\nu} F_{\mu \nu} \,,
\label{eq:LIaD}
\end{align}
where $\psi$ is a four-component spinor for the Dirac DM particle.~If $g_1$ and $\Lambda$ are the same as in Eq.~(\ref{eq:LIa}), the non-relativistic amplitude for DM-electron scattering predicted by Eq.~(\ref{eq:LIaD}) coincides with the one in Eq.~(\ref{eq:MNRa}), as shown in App.~\ref{sec:amplitudes}.~Let us now focus on the magnetic dipole coupling between DM and the photon.~The interaction Lagrangian associated with this electromagnetic moment reads
\begin{align}
\mathscr{L}_m = \frac{g_2}{\Lambda} \, \bar{\psi} \sigma^{\mu \nu} \psi \, F_{\mu \nu}\,,
\label{eq:LIm}
\end{align}
where $g_2$ is a dimensionless coupling constant and $\Lambda$ a mass scale which can in principle be different from the one introduced in Sec.~(\ref{eq:LIa}), but which for simplicity we assume to coincide with the former.~Any difference between the two mass scales can be reabsorbed in a redefinition of $g_2$.~In the non-relativistic limit, Eq.~(\ref{eq:LIm}) generates the amplitude for DM-electron scattering,
\begin{align}
\mathcal{M}&=\frac{e g_2}{\Lambda}  \Bigg\{
    4m_e\delta^{s's}\delta^{r'r} +\frac{16m_\chi m_e}{|\mathbf{q}|^2}  i\mathbf{q} \cdot \left(\mathbf{v}_{\rm el}^\perp \times \xi^{s' \dagger} \mathbf{S}_\chi \xi^s \right)\delta^{r'r} 
    \nonumber\\
    &-  \frac{8 g_em_\chi}{|\mathbf{q}|^2} \Bigg[\left( \mathbf{q} \cdot \xi^{s'\dagger} \mathbf{S}_\chi \xi^s \right)\left( \mathbf{q} \cdot \eta^{r'\dagger } \mathbf{S}_e \eta^r \right)
    - |\mathbf{q}|^2  \left( \xi^{s'\dagger} \mathbf{S}_\chi \xi^s \right)\cdot \left( \eta^{r'\dagger } \mathbf{S}_e \eta^r \right)
    \Bigg] \Bigg\}\,.
\label{eq:MNRm}
\end{align}
Finally, the interaction Lagrangian associated with the electric dipole moment can be written as follows
\begin{align}
\mathscr{L}_e = \frac{g_3}{\Lambda} \, i \bar{\psi} \sigma^{\mu \nu} \gamma^5 \psi \, F_{\mu \nu}\,,
\label{eq:LIe}
\end{align}
where $g_3$ is a dimensionless coupling constant and $\Lambda$ a mass scale (see comment on $\Lambda$ below Eq.~(\ref{eq:LIm})).~In the non-relativistic limit, Eq.~(\ref{eq:LIe}) generates the amplitude for DM-electron scattering,
\begin{align}
\mathcal{M}&= \frac{e g_3}{\Lambda} \frac{16 m_\chi m_e}{|\mathbf{q}|^2} i\mathbf{q} \cdot \left( \xi^{s'\dagger} \mathbf{S}_\chi \xi^s \right)\delta^{r' r} \,.
\label{eq:MNRe}
\end{align}
To summarise:~we refer to ``Dirac DM hypothesis'' as a scenario where the full interaction Lagrangian, $\mathscr{L}_I^{(D)}$, is given by
\begin{align}
\mathscr{L}_I^{(D)} = \mathscr{L}^{(D)}_a + \mathscr{L}_m +\mathscr{L}_e \, ,
\end{align}
and the amplitude for DM-electron scattering is given by the sum of Eqs.~(\ref{eq:MNRa}), (\ref{eq:MNRm}) and (\ref{eq:MNRe}).~We refer to ``Majorana DM hypothesis'' as a scenario with an interaction Lagrangian given in Eq.~\eqref{eq:LIa} such that the amplitude for DM-electron scattering is equal to Eq.~(\ref{eq:MNRa}).

\section{Dark matter direct detection via atomic ionisations}
\label{sec:ionisation}
To compare the predictions of the models introduced in the previous section with observations performed at DM direct detection experiments using xenon as a target material, we calculate the rate of DM-induced transitions from an initial electron state $|\mathbf{e}_1\rangle$ to a final electron state $|\mathbf{e}_2\rangle$~\cite{Catena:2019gfa},
\begin{align}
\mathscr{R}_{1\rightarrow 2}&=\frac{n_{\chi}}{16 m^2_{\chi} m^2_e}
\int \frac{{\rm d}^3 q}{(2 \pi)^3} \int {\rm d}^3 v f_{\chi}(\mathbf{v}) (2\pi) \delta(E_f-E_i) \overline{\left| \mathcal{M}_{1\rightarrow 2}\right|^2}\,,
\label{eq:transition rate}
\end{align}
where $\rho_\chi=0.4$~GeV/cm$^3$~\cite{Catena:2009mf}, $n_\chi=\rho_\chi/m_\chi$ is the local DM number density, and $f_\chi(\mathbf{v})$ is the local DM velocity distribution.~For $f_\chi(\mathbf{v})$, we assume a Maxwell-Boltzmann~distribution truncated at the escape velocity $v_{\rm esc}=544$~km~s$^{-1}$~\cite{Smith:2006ym} and boosted to the detector rest frame~\cite{Lewin:1995rx},
\begin{align}
    f_\chi(\mathbf{v})&= \frac{1}{N_{\rm esc}\pi^{3/2}v_0^3}\exp\left[-\frac{(\mathbf{v}+\mathbf{v}_\oplus)^2}{v_0^2} \right]
    \times\Theta\left(v_{\rm esc}-|\mathbf{v}+\mathbf{v}_\oplus|\right)\, ,
    \label{eq:fv}
\end{align}
where $v_0=220$~km~s$^{-1}$ is the most probable speed~\cite{Kerr:1986hz}, $\mathbf{v}_\oplus$, with~$\left|\mathbf{v}_\oplus\right|=244$~km~s$^{-1}$, the detector velocity in the galactic rest frame~\cite{McCabe:2013kea}, and $N_{\rm esc}$ a normalisation constant,
\begin{align}
N_{\rm esc}\equiv \erf\left(\frac{v_{\rm esc}}{v_0}\right)-\frac{2}{\sqrt{\pi}} \frac{v_{\rm esc}}{v_0} \exp\left(-\frac{v_{\rm esc}^2}{v_0^2}\right) \,.
\end{align}
A central element in Eq.~(\ref{eq:transition rate}) is the squared electron transition amplitude~\cite{Catena:2019gfa}, 
\begin{align}
    \overline{\left| \mathcal{M}_{1\rightarrow 2}\right|^2}\equiv \overline{\left|\int  \frac{{\rm d}^3 k}{(2 \pi)^3} \, \psi_2^*(\mathbf{k}+\mathbf{q})  
\mathcal{M}(\mathbf{q},\mathbf{v}_{\rm el}^\perp)
\psi_1(\mathbf{k}) \right|^2}\,, 
\label{eq:transition amplitude}
\end{align}
where $\mathcal{M}$ is the amplitude for DM scattering by free electrons while $\psi_1$ and $\psi_2$ are the initial and final state electron wave functions, respectively.~In Eq.~(\ref{eq:transition amplitude}), a bar denotes an average (sum) over initial (final) spin states, $\mathbf{q}=\mathbf{p}-\mathbf{p}'$ is the momentum transfer, while $\mathbf{p}$ and $\mathbf{p}'$ ($\mathbf{k}$ and $\mathbf{k}'$ ) are the initial and final DM particle (electron) momenta, respectively.~Furthermore, we introduce
\begin{align}
\mathbf{v}_{\rm el}^\perp &= \frac{\left( \mathbf{p} + \mathbf{p}' \right)}{2 m_{\chi}} - \frac{\left( \mathbf{k} + \mathbf{k}' \right)}{2 m_e}  =\mathbf{v} - \frac{\mathbf{q}}{2\mu_{\chi e}} - \frac{\mathbf{k}}{m_e}\, ,
\end{align}
where $\mu_{\chi e}$ is the reduced DM-electron mass and $\mathbf{v}\equiv \mathbf{p}/m_\chi$ the incoming DM particle velocity~\footnote{If the DM-electron scattering were elastic, $\mathbf{v}_{\rm el}^\perp\cdot \mathbf{q}=0$ would apply, which justifies the notation adopted here.}.~Finally, the initial and final state energies in the delta function in Eq.~(\ref{eq:transition rate}) are given by
\begin{align}
    E_i &= m_\chi + m_e + \frac{m_\chi}{2}v^2 + E_1\, , \label{eq: energy initial}\\
    E_f &= m_\chi + m_e + \frac{|m_\chi\mathbf{v}-\mathbf{q}|^2}{2m_\chi} + E_2\,, \label{eq: energy final}
\end{align}
where we denote by $E_1$ and $E_2$ the electron initial and final energies, and by $\Delta E_{1\rightarrow 2}=E_2-E_1$ their difference.

The electron initial state, $|\mathbf{e}_1\rangle$, in Eq.~(\ref{eq:transition rate}) is a bound state characterised by the principal, angular and magnetic quantum numbers~$(n,\ell,m)$, respectively.~At large distances from the target atom, the final state $|\mathbf{e}_2\rangle$ describes a free particle.~As a result, it can be defined in terms of the quantum numbers $(k^\prime,\ell^\prime,m^\prime)$, where~$k^\prime$ is the electron momentum at infinitely large distances from the atom, while $\ell^\prime$ and $m^\prime$ are its angular and magnetic quantum numbers.~We can now express the differential ionisation rate~${\rm d}\mathscr{R}_{\rm ion}^{n\ell}/{\rm d}\ln E_e$, of a full $(n,\ell)$ atomic orbital as follows
\begin{align}
    \frac{\dd\mathscr{R}_{\rm ion}^{n\ell}}{\dd\ln E_e}&=
    \sum_{m=-\ell}^\ell \sum_{\ell^\prime=0}^\infty\sum_{m^\prime=-\ell^\prime}^{\ell^\prime} \frac{V k^{\prime 3}}{(2\pi)^3} \mathscr{R}_{1\rightarrow 2} \nonumber\\
&=\frac{n_{\chi}}{128 \pi \,m^2_{\chi} m^2_e}
\int \dd q \; q \int \frac{{\rm d}^3 v}{v} \,f_{\chi}(\mathbf{v}) \Theta(v-v_{\rm min}) \overline{\left| \mathcal{M}^{n\ell}_{\rm ion}\right|^2}\,, 
\label{eq:ionization spectrum}
\end{align}
where $E_e=k'^{2}/(2m_e)$ is the energy carried by the ejected electron, $\Theta(x)$ is the step-function, and $v_{\rm min}=\Delta E_{1\rightarrow 2}/q + q/(2 m_\chi)$.~Here, the squared ionisation amplitude $\overline{\left| \mathcal{M}^{n\ell}_{\rm ion}\right|^2}$ is defined as~\cite{Catena:2019gfa}
\begin{align}
    \overline{\left| \mathcal{M}^{n\ell}_{\rm ion}\right|^2} &\equiv V \frac{4k^{\prime 3}}{(2\pi)^3} \sum_{m = -\ell}^\ell \sum_{\ell^\prime=0}^\infty \sum_{m^\prime = -\ell^\prime}^{\ell^\prime} \overline{\left| \mathcal{M}_{1\rightarrow 2}\right|^2}\, , 
\label{eq:ionization amplitude}
\end{align}
and  admits a general decomposition in terms of atomic ($W_{i}^{n\ell}$) and DM ($R^{n\ell}_i$) response functions,
\begin{align}
\label{eq:RW}
   \overline{| \mathcal{M}_{\rm ion}^{n\ell}|^2}= \sum_{i=1}^4    R^{n\ell}_i\left(\vPerpEl,\frac{\mathbf{q}}{m_e}\right) W_{i}^{n\ell}(k^\prime,\mathbf{q})\, .
\end{align}
As a function of scalar and vectorial form factors~\cite{Catena:2019gfa},
\begin{align}
	f_{1\rightarrow 2}(\mathbf{q}) &= \int\frac{\dd^3 k}{(2\pi)^3}\psi^*_{2}(\mathbf{k}+\mathbf{q})\psi_{1}(\mathbf{k})\, , 
	\nonumber\\ 
	\mathbf{f}_{1\rightarrow 2}(\mathbf{q})&= \int \frac{\dd^3 k}{(2\pi)^3}\psi^*_2(\mathbf{k}+\mathbf{q})\,\left(\frac{\mathbf{k}}{m_e}\right)\,\psi_1(\mathbf{k})\, ,
\end{align}
the four atomic response functions, $W_{i}^{n\ell}$, $i=1,\dots,4$, appearing in Eq.~(\ref{eq:RW}) can be written as follows~\cite{Catena:2019gfa}
\begin{subequations}
\label{eq: atomic responses}
\begin{align}
    W_{1}^{n\ell}(k^\prime,\mathbf{q}) &\equiv V\frac{4k^{\prime 3}}{(2\pi)^3} \sum_{m = -\ell}^\ell \sum_{\ell^\prime=0}^\infty \sum_{m^\prime = -\ell^\prime}^{\ell^\prime} \left|f_{1\rightarrow 2}(q)\right|^2 \, ,\label{eq: atomic response 1}\\
    W_{2}^{n\ell}(k^\prime,\mathbf{q}) &\equiv V\frac{4k^{\prime 3}}{(2\pi)^3} 
    \sum_{m = -\ell}^\ell \sum_{\ell^\prime=0}^\infty \sum_{m^\prime = -\ell^\prime}^{\ell^\prime}\frac{\mathbf{q}}{m_e}\cdot f_{1\rightarrow 2}(\mathbf{q})\mathbf{f}^{\,*}_{1\rightarrow 2}(\mathbf{q}) \, ,\label{eq: atomic response 2}\\
    W_{3}^{n\ell}(k^\prime,\mathbf{q}) &\equiv V\frac{4k^{\prime 3}}{(2\pi)^3} \sum_{m = -\ell}^\ell \sum_{\ell^\prime=0}^\infty \sum_{m^\prime = -\ell^\prime}^{\ell^\prime} |\mathbf{f}_{1\rightarrow 2}(\mathbf{q})|^2\, ,\label{eq: atomic response 3}\\
    W_{4}^{n\ell}(k^\prime,\mathbf{q}) &\equiv V\frac{4k^{\prime 3}}{(2\pi)^3} 
        \sum_{m = -\ell}^\ell \sum_{\ell  ^\prime=0}^\infty \sum_{m^\prime = -\ell^\prime}^{\ell^\prime} \left|\frac{\mathbf{q}}{m_e}\cdot\mathbf{f}_{1\rightarrow 2}(\mathbf{q})\right|^2\, ,\label{eq: atomic response 4}
\end{align}
\end{subequations}
where $V=(2\pi)^3 \delta^{(3)}(0)$.~The first one, $W_{1}^{n\ell}$, is the standard ``ionisation form factor'' common in the light DM literature.~The remaining three atomic response functions $W_{j}^{n\ell}$, $j=2,3,4$ were identified in~\cite{Catena:2019gfa} for the first time and computed with the \textit{DarkARC} tool~\cite{Emken2019}. They describe distortions in the ionisation spectrum induced by the finite dispersion of the initial state electron momentum distribution.~Notice that all atomic responses in Eq.~(\ref{eq:}) are required to investigate the electromagnetically interacting DM models presented here in a self-consistent manner.
In~\cite{Catena:2019gfa}, the xenon targets were modeled as isolated atoms, and wave function distortions due to the condensed phase of the liquid xenon are not taken into account.
This approximation renders the prediction of ionisation events conservative, as the broadening of the electron energy levels into energy bands effectively reduces the energy gap which in turn enhances the DM~induced ionisation rate.
For example, while we assume a binding energy of~$\sim12.4$~eV for the 5p orbital of xenon, the band gap of liquid xenon is actually closer to 9.2~eV~\cite{Asaf1974}.

The four DM response functions are model dependent and in the case of Dirac DM they depend on couplings and momenta as follows
\begin{subequations}
\label{eq: DM response functions}
\begin{align}
   R^{n\ell}_1\left(\vPerpEl,\frac{\mathbf{q}}{m_e}\right) &\equiv c_1^2 + \frac{j_\chi(j_\chi+1)}{12}\Bigg\{ 3c_4^2  + 4c_8^2(\vPerpEl)^2 +(2c_9^2+4c_{11}^2+2c_4c_6)\left(\frac{\mathbf{q}}{m_e}\right)^2  \nonumber \\
   &+c_6^2 \left(\frac{\mathbf{q}}{m_e}\right)^4+4c_5^2\left[\left(\frac{\mathbf{q}}{m_e}\right)^2(\vPerpEl)^2 - \left(\vPerpEl\cdot\frac{\mathbf{q}}{m_e}\right)^2 \right] \Bigg\}\, ,\label{eq: DM response 1}\\
R^{n\ell}_2\left(\vPerpEl,\frac{\mathbf{q}}{m_e}\right) &\equiv-4c_8^2\left(\frac{\mathbf{q}}{m_e}\cdot \vPerpEl\right)\left[\frac{j_\chi(j_\chi+1)}{6}\left(\frac{\mathbf{q}}{m_e}\right)^{-2}\right]\, ,\label{eq: DM response 2}\\
   R^{n\ell}_3\left(\vPerpEl,\frac{\mathbf{q}}{m_e}\right) &\equiv\frac{j_\chi(j_\chi+1)}{12}\left[4c_8^2+4c_5^2\left(\frac{\mathbf{q}}{m_e}\right)^{2}\right] \, ,\label{eq: DM response 3}\\
   R^{n\ell}_4\left(\vPerpEl,\frac{\mathbf{q}}{m_e}\right) &\equiv -c_5^2 \frac{j_\chi(j_\chi+1)}{3} \label{eq: DM response 4}\, ,
\end{align}
\end{subequations}
where $j_\chi=1/2$ and 
\begin{align}
c_1 &= \frac{4 e g_2 m_e}{\Lambda},  &c_5 &= \frac{16 e  g_2 m_\chi m^2_e}{\Lambda |\mathbf{q}|^2}, &c_8 &= \frac{8 e g_1 m_e m_\chi}{\Lambda^2}, &c_{11} &= \frac{16 e g_3 m_\chi m_e^2}{\Lambda |\mathbf{q}|^2},    \nonumber\\
c_4 &= \frac{8 e g_e g_2 m_\chi}{\Lambda}, &c_6 &= -\frac{8 e g_e g_2 m_\chi m_e^2}{\Lambda |\mathbf{q}|^2},  &c_9 &= -\frac{4 e g_e g_1 m_e m_\chi}{\Lambda^2}. 
\end{align}
The only interference term arising in this case is the one proportional to $c_4 c_6$ (see the definition of $R^{n\ell}_1$ in Eq.~(\ref{eq: DM response 1}), first line).~Notice that this interference arises within the magnetic dipole interaction and it is not an interference between different electromagnetic moments.~In the case of Majorana DM, the fourth DM response, $R^{n\ell}_4$, is identically zero.~The remaining three DM response functions keep the same definition as above, but now with coupling constants given by
\begin{align}
c_8 &= \frac{8 e g_1 m_e m_\chi}{\Lambda^2}\,, & c_9 &= -\frac{4 e g_e g_1 m_e m_\chi}{\Lambda^2} \,, 
\end{align}
and $c_1=c_4=c_5=c_6=c_{11}=0$.~We refer to~\cite{Catena:2019gfa} for further details on the numerical evaluation of the ionisation rate in Eq.~(\ref{eq:ionization spectrum}) and on our choice for the electron wave functions~$\psi_1$ and~$\psi_2$.

\section{Hypothesis testing}
\label{sec:stat}
In this section, we introduce the statistical methods that we use to compare the Dirac DM hypothesis ($\mathscr{H}_{D}$) with the Majorana DM hypothesis ($\mathscr{H}_{M}$).~The $\mathscr{H}_{D}$ hypothesis is defined by $g_1^2\ge0$, $g_2^2\ge0$ and $g_{3}^2\ge0$, whereas the $\mathscr{H}_{M}$ hypothesis corresponds to $g_1^2\ge0$, $g_2^2=0$ and $g_{3}^2=0$.~The two hypotheses are therefore nested, which implies $\mathscr{H}_{D} \rightarrow \mathscr{H}_{M}$ in the $g_2^2=g_3^2=0$ limit.~We compare the $\mathscr{H}_{D}$ hypothesis with the $\mathscr{H}_{M}$ hypothesis by means of the log-likelihood ratio as a test statistic~\cite{Cowan:2010js},
\begin{align}
t = -2 \ln \frac{\max_{\boldsymbol{\Theta}\in\Omega_M}\mathscr{L}(\mathscr{D}|\boldsymbol{\Theta})}{\max_{\boldsymbol{\Theta}\in\Omega_D}\mathscr{L}(\mathscr{D}|\boldsymbol{\Theta})} \,,
\label{eq:q}
\end{align}
where $\boldsymbol{\Theta}=\{\theta_1\equiv g_1^2/\Lambda^4, \theta_2\equiv g_2^2/\Lambda^2, \theta_3\equiv g_3^2/\Lambda^2\}$, $\Omega_M=\{\boldsymbol{\Theta}~:~\theta_1\ge 0,  \theta_2=0, \theta_3=0 \}$, $\Omega_D=\{\boldsymbol{\Theta}~:~\theta_1\ge0,  \theta_2\ge0, \theta_3\ge0 \}$ and $\mathscr{D}$ is a dataset.~In our case, $\mathscr{D}$ consists of $(16-n_{\rm th})$ independent observations, i.e.~$\mathscr{D}=\{\mathscr{N}_{n_e}\}$, where $\mathscr{N}_{n_e}$, is the number of DM-induced atomic ionisations producing $n_e=n_{\rm th},\dots,15$ observable electrons in the detector.~Here, $n_{\rm th}$ is the experimental threshold, i.e.~the minimum number of observable electrons per ionisation.~Since DM has so far escaped detection, we generate the dataset $\mathscr{D}$ via Monte Carlo~(MC) simulations.~Specifically, we sample $\mathscr{D}$ from a benchmark point $\boldsymbol{\Theta}'$ in parameter space.~We will consider two choices for  $\boldsymbol{\Theta}'$:
\begin{align}
\boldsymbol{\Theta}'&=\left\{ g_1^2/\Lambda^4=\mathcal{C},  g_2^2/\Lambda^2=0, g_3^2/\Lambda^2=0 \right\} \,,\label{eq:benchmark M} \\
\boldsymbol{\Theta}'&=\left\{ g_1^2/\Lambda^4=\mathcal{C}_1,  g_2^2/\Lambda^2=\mathcal{C}_2, g_3^2/\Lambda^2=\mathcal{C}_3 \right\} \, . \label{eq:benchmark D}
\end{align}
In the former case we generate $\mathscr{D}$ under the Majorana hypothesis, in the latter one we sample $\mathscr{D}$ under the alternative, Dirac hypothesis (we determine $\mathcal{C}$, $\mathcal{C}_1$, $\mathcal{C}_2$ and $\mathcal{C}_3$ in Sec.~\ref{sec:results}).~For each bin, we assume a Poisson likelihood, 
\begin{equation}
\mathscr{L}(\mathscr{D}|\boldsymbol{\Theta}) = \prod_{n_e=n_{\rm th}}^{15} \frac{\left(\mathscr{B}_{n_e}+\mathscr{S}_{n_e}(\boldsymbol{\Theta})\right)^{\mathscr{N}_{n_e}}}{\mathscr{N}_{n_e}!} e^{-\left(\mathscr{B}_{n_e}+\mathscr{S}_{n_e}(\boldsymbol{\Theta})\right)} \,,
\end{equation}
where 
\begin{equation}
\mathscr{S}_{n_e}(\boldsymbol{\Theta}) = \mathcal{E} \sum_{(n,\ell)\in \mathscr{A}} \int {\rm d} E_e \, \mathcal{P}(n_e|E_e) \, \frac{\dd\mathscr{R}_{\rm ion}^{n\ell}}{\dd E_e} \,,
\label{eq:S_i}
\end{equation}
$ \mathcal{E}$ is the experimental exposure and $\mathcal{P}(n_e|E_e)$ is the probability of producing $n_e$ observable electrons in an atomic ionisation when the energy carried by the primary electron is $E_e$~\cite{Essig:2012yx,Essig:2017kqs}.~In the definition of $\mathscr{S}_{n_e}$, we sum over the five outermost occupied orbitals for xenon, $\mathscr{A}\equiv \{$4s, 4p, 4d, 5s, 5p$\}$.~Here, $\mathscr{B}_{n_e}$ is the number of background events producing $n_e$ observable electrons.~Since $\mathscr{S}_{n_e}$ is expected to be significantly larger than $\mathscr{B}_{n_e}$ when $\mathscr{H}_D$ can be discriminated from $\mathscr{H}_M$, we can safely neglect the experimental background contribution to the likelihood function and set $\mathscr{B}_{n_e}=0$ for the purposes of this study.  

By repeatedly sampling $\mathscr{D}$ under $\mathscr{H}_M$, one obtains the probability density function of $t$ under $\mathscr{H}_M$, i.e.~$f(t|\mathscr{H}_M)$.~Similarly, by repeatedly simulating $\mathscr{D}$ under $\mathscr{H}_D$, one obtains the probability density function of $t$ under $\mathscr{H}_D$, which we denote by $f(t|\mathscr{H}_D)$.~The significance for rejecting the Majorana DM hypothesis in favour of the Dirac DM hypothesis, $\mathcal{Z}$, is then given by
\begin{equation}
\mathcal{Z} = \Phi^{-1}(1-p)\,,
\label{eq:Z}
\end{equation}
where $\Phi$ is the cumulative distribution function of a Gaussian probability density of variance 1 and mean 0, whereas
\begin{equation}
p = \int^{\infty}_{t_{\rm med}} {\rm d}t\, f(t|\mathscr{H}_M)\,,
\label{eq:pvalue}
\end{equation}
is the $p$-value for rejecting $\mathscr{H}_M$ in favour of $\mathscr{H}_D$.~Here, $t_{\rm med}$ is the median of the probability density function $f(t|\mathscr{H}_D)$.~In all numerical applications, we obtain $t_{\rm med}$ by random sampling $10^4$ values for $t$ under $\mathscr{H}_D$ and computing the median of this discrete sample.~On the other hand, the method we use to estimate the integral of $f(t|\mathscr{H}_M)$ above $t_{\rm med}$ depends on whether $t_{\rm med}$ lies in the tail of $f(t|\mathscr{H}_M)$ or not.~We compute the integral in Eq.~(\ref{eq:pvalue}) via MC integration, i.e. by random sampling $10^5$ values for $t$ under $\mathscr{H}_M$ and then counting the relative fraction of them above $t_{\rm med}$, if at least 100 sampled values for $t$ lie above $t_{\rm med}$.~We switch to an asymptotic expression for $f(t|\mathscr{H}_M)$ and do the integral in Eq.~(\ref{eq:pvalue}) numerically, when within the $10^5$ sampled values for $t$ under $\mathscr{H}_M$ less then 100 lie above $t_{\rm med}$.~We obtain such an asymptotic expression for $f(t|\mathscr{H}_M)$ as explained in detail below.~In the large-sample limit, $f(t|\mathscr{H}_M)$ is approximated by~\cite{Kopylev2011May}
\begin{align}
t \sim \min_{\boldsymbol{\Theta}\in\Omega_M} Q(\boldsymbol{\Theta}|\hat{\boldsymbol{\Theta}}) - \min_{\boldsymbol{\Theta}\in\Omega_D} Q(\boldsymbol{\Theta}|\hat{\boldsymbol{\Theta}}) \,,
\label{eq:Q}
\end{align}
where $Q(\boldsymbol{\Theta}|\hat{\boldsymbol{\Theta}})=(\boldsymbol{\Theta}-\hat{\boldsymbol{\Theta}})^T \mathcal{I}(\boldsymbol{\Theta}')(\boldsymbol{\Theta}-\hat{\boldsymbol{\Theta}})$, $\mathcal{I}(\boldsymbol{\Theta}')$ is the Fisher information matrix at the benchmark point $\boldsymbol{\Theta}'$\footnote{Strictly speaking, $\mathcal{I}(\boldsymbol{\Theta})$ should be evaluated at the maximum likelihood estimator, $\bar{\boldsymbol{\Theta}}$, but $\bar{\boldsymbol{\Theta}}=\boldsymbol{\Theta}'$ under the assumption of Asimov data~\cite{Cowan:2010js}.} and the stochastic variable $\hat{\boldsymbol{\Theta}}=\{\hat{\theta}_1,\hat{\theta}_2,\hat{\theta}_3 \}$ follows a multivariate Gaussian distribution of mean $\boldsymbol{\Theta}'$ and covariance matrix $\mathcal{I}^{-1}(\boldsymbol{\Theta}')$.~Eq.~(\ref{eq:Q}), with the quadratic form $Q$ given above, assumes that $\mathcal{I}$ is positive definite.~This explains our choice of defining $\boldsymbol{\Theta}$ in terms of squared coupling constants.~Indeed, if we defined $\boldsymbol{\Theta}\equiv\{ g_1/\Lambda^2,g_2/\Lambda,g_2/\Lambda \}$, the Fisher matrix would be singular at $\{g_1/\Lambda^2=\sqrt{\mathcal{C}}, g_2/\Lambda=0, g_3/\Lambda=0 \}$.~Let us now use Eq.~(\ref{eq:Q}) to find an expression for $f(t|\mathscr{H}_M)$ that is valid in the large-sample limit.~If we were interested in testing the ``null'' hypothesis defined by $\boldsymbol{\Theta}'=(\alpha, \beta, \gamma)$ with $\alpha$, $\beta$, and $\gamma$ strictly positive real numbers and $\Omega_M=\{\boldsymbol{\Theta}~:~\theta_1\ge 0,  \theta_2=\beta, \theta_3=\gamma \}$ against an alternative hypothesis characterised by $\Omega_D=\{\boldsymbol{\Theta}~:~\theta_1\ge 0,  \theta_2\ge 0, \theta_3\ge 0 \}$, then the benchmark point $\boldsymbol{\Theta}'$ would be an interior point of the set $\Omega_D$ and
\begin{align}
\min_{\boldsymbol{\Theta}\in\Omega_D} Q(\boldsymbol{\Theta}|\hat{\boldsymbol{\Theta}})&= 0\,, \nonumber\\
\min_{\boldsymbol{\Theta}\in\Omega_M} Q(\boldsymbol{\Theta}|\hat{\boldsymbol{\Theta}})&=(\hat{\theta}_2-\beta,\hat{\theta}_3-\gamma) \mathcal{I}_{2\times2}(\boldsymbol{\Theta}')(\hat{\theta}_2-\beta,\hat{\theta}_3-\gamma)^T \,,
\end{align}
where $\mathcal{I}_{2\times2}(\boldsymbol{\Theta}')$ is $\mathcal{I}(\boldsymbol{\Theta}')$  restricted to the 2-dimensional space spanned by $(\theta_2,\theta_3)$.~In this example, the probability density function of $(\hat{\theta}_2-\beta,\hat{\theta}_3-\gamma)$ is a multivariate Gaussian of mean $(\beta,\gamma)$ and covariance matrix $\mathcal{I}^{-1}_{2\times2}(\boldsymbol{\Theta}')$, consistently with our definition of $\hat{\boldsymbol{\Theta}}$.~Consequently, $t$ would in this case follow a chi-square distribution with 2 degrees of freedom.~However, when the benchmark point $\boldsymbol{\Theta}'$ lies on the boundary of $\Omega_D$, as when sampling $\mathscr{D}$ under $\mathscr{H}_{M}$ and $\boldsymbol{\Theta}'=\{ g_1^2/\Lambda^4=\mathcal{C},  g_2^2/\Lambda^2=0, g_3^2/\Lambda^2=0 \}$, the asymptotic distribution of $t$, $\bar{\chi}^2$, is a chi-bar-square distribution, i.e.~a linear combination of chi-square distributions of different degrees of freedom~\cite{Shapiro1985Apr}. Consequently,
\begin{align}
\mathscr{P}(\bar{\chi}^2\le x) = \sum_{i=0}^{n_p} w_i\, \mathscr{P}(\chi^2_i \le x) \,,
\label{eq:wi}
\end{align}
where $\mathscr{P}(\bar{\chi}^2\le x)$ and $\mathscr{P}(\chi^2_i \le x$) are the cumulative probability density functions at $x\ge 0$ of $\bar{\chi}^2$ and a chi-square distribution with $i$ degrees of freedom, respectively.~Here, $n_p$ is the number of model parameters ($n_p=3$ in our case) and $\chi_0^2$ a one-dimensional Dirac delta.~The weights $w_i$ in Eq.~(\ref{eq:wi}) depend on $\Omega_M$, $\Omega_D$ and $\mathcal{I}$ and are analytically known only in a few, specific cases.~For example, when $\Omega_M=\{\boldsymbol{\Theta}~:~\theta_1\ge 0,  \theta_2=0 \}$, $\Omega_D=\{\boldsymbol{\Theta}~:~\theta_1\ge0,  \theta_2\ge0 \}$, i.e. in the $n_p=2$ case, one finds $w_0=1/2 - \sin^{-1}(\rho)/(2\pi)$, $w_1=1/2$ and $w_2=\sin^{-1}(\rho)/(2\pi)$, where $\rho$ is the correlation coefficient between $\theta_1$ and $\theta_2$~\cite{Kopylev2011May}.~This can be extracted analytically from the inverse Fisher matrix, $\mathcal{I}^{-1}(\boldsymbol{\Theta}')$.~To the best of our knowledge, an analytic expression for $w_i$ is not available in our case\footnote{The $n_p=3$ case is treated in~\cite{Shapiro1985Apr} with a different definition for $\Omega_M$ and in~\cite{Kopylev2011May} without arriving at an analytic expression for $w_i$.}.~We therefore estimate the weights $w_i$ by numerically solving the linear system
\begin{subequations}
\label{eq:system}
\begin{align}
\sum_{i=0}^{n_p} w_i&=1\, ,\\
w_0+w_2&=1/2 \, ,\\
\mathscr{P}(\bar{\chi}^2\le x_{a})_{\rm MC} &= \sum_{i=0}^{n_p} w_i\, \mathscr{P}(\chi^2_i \le x_a)\, ,\\
\mathscr{P}(\bar{\chi}^2\le x_{b})_{\rm MC} &= \sum_{i=0}^{n_p} w_i\, \mathscr{P}(\chi^2_i \le x_b) \,,
\end{align}
\end{subequations}
where the first two equations correspond to constraints that the weights $w_i$ must fulfil in general~\cite{Shapiro1985Apr}, $x_a$ and $x_b$ are arbitrary real numbers that we set to 0 and 1, and $\mathscr{P}(\bar{\chi}^2\le x)_{\rm MC}$ is the cumulative distribution function of $\bar{\chi}^2$ at $x=x_a$ or $x=x_b$ that we extract from $10^5$ random samples of $t$ under $\mathscr{H}_M$.
The choice of~$x_a=0$ is convenient, since only the first term in Eq.~\eqref{eq:wi}, being a Dirac delta, contributes to the probability of vanishing $t$~values, and counting the instances of $t=0$ in the sample directly yields the first weight~$w_0$.
By solving Eq.~(\ref{eq:system}), we obtain an analytic estimate for $f(t|\mathscr{H}_M)$ which enables us to compute the number of signal events required to reject the Majorana hypothesis with an arbitrarily large statistical significance.~Here, we restrict ourselves to $\mathcal{Z}\le 6$.

\section{Numerical results}
\label{sec:results}
In order to discriminate the case of Majorana~DM, our null hypothesis~$\mathscr{H}_M$, from the scenario where DM is made of Dirac~fermions, our alternative hypothesis~$\mathscr{H}_D$, we need to compare a one-parameter model (with the squared anapole coupling~$g_1^2$) to a model with three parameters (the squared couplings~$g_1^2,g_2^2,g_3^2$ for anapole, magnetic dipole, and electric dipole respectively).
Equipped with a three-dimensional parameter space, the Dirac model allows for a vast number of phenomenologically very different configurations.
Therefore, we need to specify the Dirac hypothesis further to allow quantitative statements about the expected significance to reject the null hypothesis in its favour.
For this purpose, we assume a number of representative hierarchies between the contributions of the anapole, magnetic dipole, and electric dipole interactions to the predicted signal event rates.
This is simplified by the absence of interferences between the three interactions.
In the following list, we define a number of benchmark hierarchies between these contributions, loosely inspired by Plato's regimes of governments.
\begin{itemize}
	\item \textbf{Democracy}(D): All three electron interactions of the Dirac DM particle contribute equally to the total signal event rate ($1$ : $1$ : $1$).
	\item \textbf{Aristocracy}(A): A sub-group of two couplings dominate the total signal event rate. For example, we label the case where the anapole and magnetic dipole dominate over the electric dipole as $\text{A}_{12}$. More specifically, the three interactions contribute to the signal rate with relative proportions ($1$ : $1$ : $10^{-3}$) in this case. We define the other two possible ``aristocratic'' hierarchies $\text{A}_{13}$ and $\text{A}_{23}$ in the same way.
	\item \textbf{Tyranny}(T): Here, the signal event rates are assumed to be dominated almost completely by a single interaction. The scenario where e.g. the magnetic dipole interaction dominates is labelled as~$\text{T}_2$ and corresponds to a relative signal contribution by anapole, magnetic dipole, and electric dipole interactions of~($10^{-3}$ : $1$ : $10^{-3}$). The hierarchies~$\text{T}_1$  and~$\text{T}_3$  are defined accordingly.
\end{itemize}
In total, these correspond to seven different versions of the Dirac hypothesis (D, $\text{A}_{12}$, $\text{A}_{13}$, $\text{A}_{23}$, $\text{T}_1$ , $\text{T}_2$ , $\text{T}_3$ ).
However, we also note that the Dirac hypothesis with~$\text{T}_{1}$ is virtually indistinguishable from the null hypothesis, and it is only kept as a consistency check.
We emphasise again that the hierarchies are not defined in terms of the (squared) couplings~$g_i^2$ but of their respective contributions to the predicted event rates.

In the Eqs.~\eqref{eq:benchmark M} and~\eqref{eq:benchmark D}, we introduced benchmark points~$\boldsymbol{\Theta}'$ in terms of the parameter(s)~$\mathcal{C}$ ($\mathcal{C}_1,\mathcal{C}_2,\mathcal{C}_3$) for the Majorana (Dirac) hypothesis.
Their values are determined by fixing the total number of signal events and choosing a Dirac hierarchy.

Furthermore, we assume that the DM~mass is known and restrict ourselves to a fixed value of~$m_\chi = 100\text{ MeV}$ throughout this statistical analysis.
Hence, we do not treat the DM~mass as an additional element or nuisance parameter of the models' parameter space~$\boldsymbol{\Theta}$, in order to emphasise the varying abilities of the different benchmark hierarchies to discriminate our two hypotheses.

\begin{figure}
\centering
\includegraphics[width=0.5\textwidth]{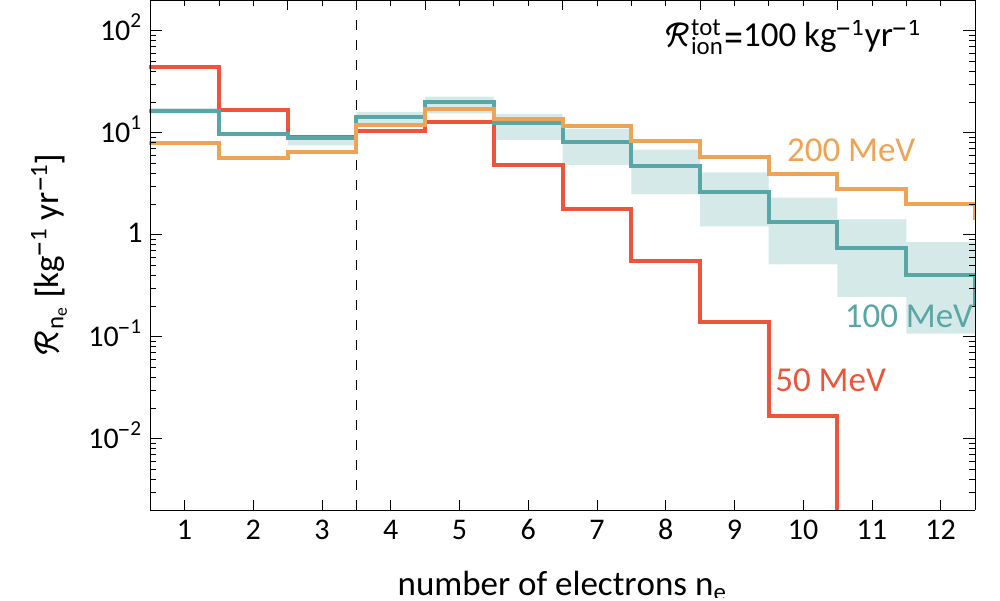}
\caption{Ionisation spectra for different DM~masses and a xenon target under the Majorana hypothesis. For each mass, the underlying coupling is fixed to yield the same total ionisation rate. The dashed, vertical line indicates the threshold~$n_{\rm th}=4$ we assume in our analysis. The blue shaded region demonstrates the impact of varying the DM~mass by 20\%.}
\label{fig: spectra mass dependence}
\end{figure}

In the scenario of an unknown mass, profiling over the mass parameter would be necessary.
In that case, we expect to require more events to reach a given statistical significance than reported in this paper.
Indeed, when varying the DM particle mass by 20\% around 100~MeV, the coupling constant $g_1/\Lambda^2$ has to change by around 10-20\% to produce the same number of signal events under the Majorana hypothesis.
In Fig.~\ref{fig: spectra mass dependence}, we illustrate the ionisation spectrum's dependency on the DM mass, and in particular the impact of varying the mass by 20\% around 100~MeV (blue shaded region).
A comparable relative variation of the individual coupling constants $g_1/\Lambda^2$, $g_2/\Lambda$ and $g_3/\Lambda$ is required to compensate for a change of 20\% in the DM mass under the Dirac hypothesis.
Consequently, we expect that the quoted numbers of signal events required to reach a given significance underestimate the values we would have obtained by profiling out the DM particle mass by approximately 20\%.
It should also be noted that for DM~masses below 100~MeV, the number of kinematically accessible electron bins varies for any mass change of a few MeV, further reducing the spectrum's degeneracy.

\begin{figure}
\centering
\includegraphics[width=0.66\textwidth]{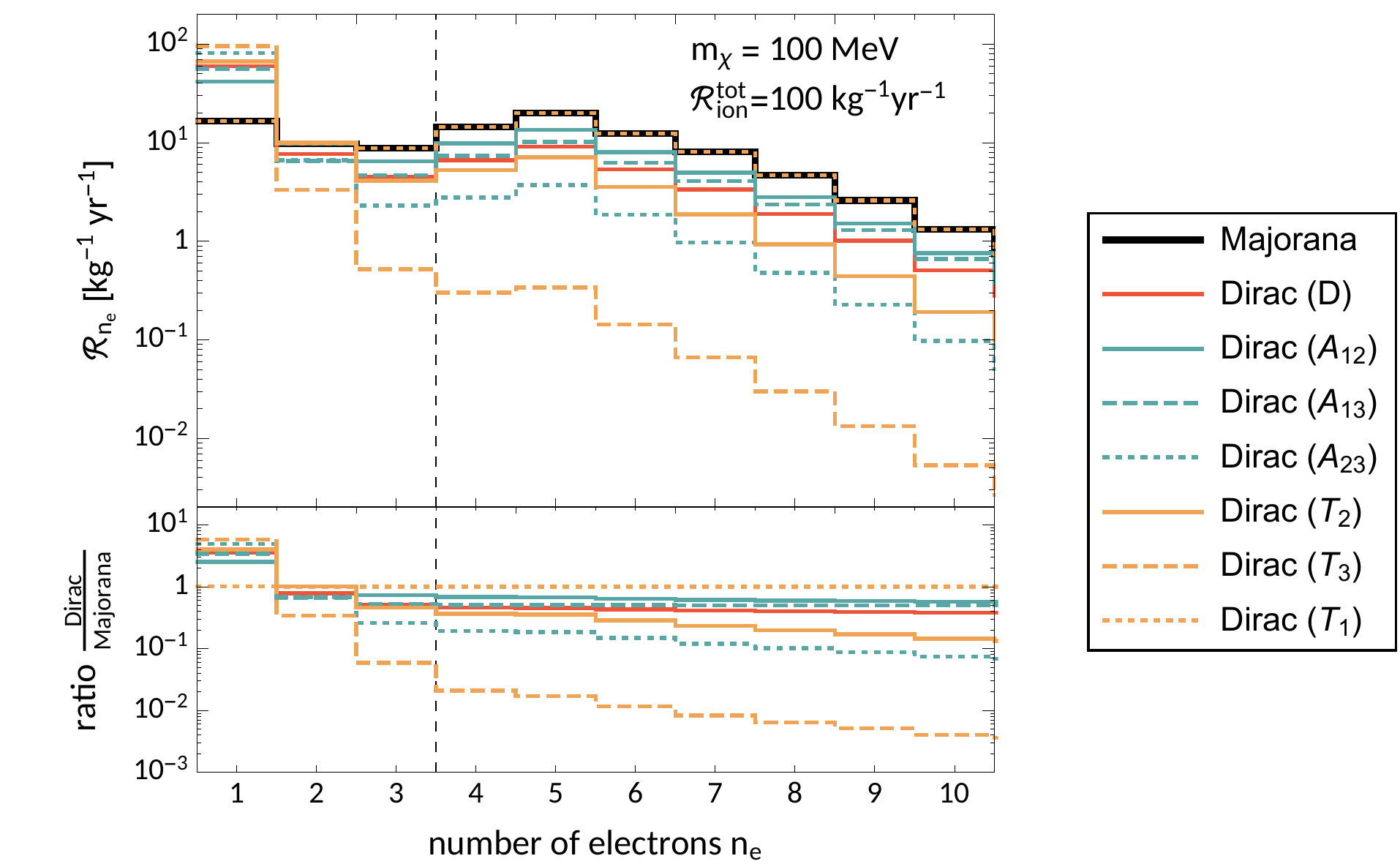}
\caption{Comparison of the spectral shapes of DM~induced ionisations of xenon between the Majorana and the different realisations of the Dirac hypothesis. The dashed, vertical line indicates the threshold~$n_{\rm th}=4$ we assume in our analysis.}
\label{fig: spectra}
\end{figure}

As a first step, we compare the xenon ionisation spectrum of the different realisations of the Dirac hypothesis~$\mathscr{H}_D$ to our null hypothesis~$\mathscr{H}_M$ of Majorana~DM.
It is defined\footnote{In the literature, this quantity is often denoted by~$\frac{\dd\mathscr{R}}{\dd n_e}$. However we choose to refrain from using this notation, as it is not the ratio of infinitesimal quantities.} as
\begin{align}
\mathscr{R}_{n_e}(\boldsymbol{\Theta}) &\equiv \frac{\mathscr{S}_{n_e}(\boldsymbol{\Theta})}{\mathcal{E}}\\
&=\sum_{(n,\ell)\in \mathscr{A}} \int {\rm d} E_e \, \mathcal{P}(n_e|E_e) \, \frac{\dd\mathscr{R}_{\rm ion}^{n\ell}}{\dd E_e}\, ,
\end{align}
where $\mathscr{S}_{n_e}(\boldsymbol{\Theta})$ was defined in Eq.~\eqref{eq:S_i}. In Fig.~\ref{fig: spectra}, the different spectra~$\mathscr{R}_{n_e}$ are shown in the case of Majorana~DM, as well as the different realisations of the Dirac hypothesis.
To demonstrate the different shapes of the spectra predicted in the different scenarios, we fix the total ionisation rate~$R_{\rm ion}^{\rm tot}\equiv \sum_{n_e = 1}^{15} \mathscr{R}_{n_e}$ to~100~$\text{ kg}^{-1}\text{ yr}^{-1}$.
The spectral deviations between the two hypotheses are the fundamental basis of the observer's ability to distinguish them in a direct detection experiment.
With this in mind, this comparison may serve as an early, qualitative indicator of which benchmark hierarchy of~$\mathscr{H}_D$ can be expected to be easier or harder to tell apart from the Majorana null hypothesis than others.

With the exception of $\text{T}_{1}$, which unsurprisingly predicts an almost identical spectrum as the null hypothesis, the Dirac~DM model generally predicts more events in the $n_e=1$ bin and fewer in the $n_e>1$ bins when compared to the Majorana spectrum.
This steeper decline of the spectrum for increasing~$n_e$ (or equivalently for increasing energy~$E_e$) originates in the additional magnetic and electric dipole moment interactions which characterise Dirac~DM.

The Dirac hierarchies involving significant signal contributions from the anapole interaction (i.e. D, $\text{A}_{12}$, and $\text{A}_{13}$) generally resemble the Majorana hypothesis more closely.
Therefore, we expect to require more events in order to be able to reject the null hypothesis in these cases.
The other hierarchies, i.e. $\text{A}_{23}$, $\text{T}_{2}$, and $\text{T}_{3}$, are associated with negligible anapole contributions and are naturally more favourable, as they give rise to an ionisation spectrum more distinct from the spectrum under the null hypothesis.
This seems to be especially true for $\text{A}_{23}$ and $\text{T}_{3}$, which have sizable contributions by electric dipole interactions.
At this point, we expect the benchmark hierarchy of $\text{T}_{3}$, i.e. Dirac~DM with dominant electric dipole interactions, to be the scenario most distinguishable from Majorana~DM.

Next, the aim is to develop a more quantitative foundation to these claims and expectations. In order to do so, we need to specify the experimental setup of the direct detection experiment, which we assume to directly detect sub-GeV DM in the hopefully not too distant future and no longer search for DM but study its properties.
We assume a xenon target with an observational threshold of~$n_{\rm th}=4$ electrons, as indicated by the dashed line in Fig.~\ref{fig: spectra}.
The choice of the experimental exposure is less crucial, as we present our results in terms of number of signals necessary to distinguish the hypotheses.
Any change of the exposure is compensated by a corresponding rescaling of the parameters~$\mathcal{C}$ and~$\mathcal{C}_i$ in Eqs.~\eqref{eq:benchmark M} and~\eqref{eq:benchmark D}.
Nonetheless, it is interesting to note what values of the effective couplings correspond to our results.
Anapole, magnetic dipole, or electric dipole interaction with a respective coupling of~$\frac{g_1}{\Lambda^2}= 4.7\cdot 10^{-3}\,\mathrm{GeV}^{-2}$, $\frac{g_2}{\Lambda}= 4.3\cdot 10^{-7}\,\mathrm{GeV}^{-1}$,  and~$\frac{g_3}{\Lambda}= 5.5\cdot 10^{-8}\,\mathrm{GeV}^{-1}$ would each lead to an expectation of $\sim 100$ DM-induced ionisation events in a xenon target detector assuming a threshold of~$n_{\rm th}=4$ electrons and an exposure of 1000 kg years.

\begin{figure}
\centering
\includegraphics[width=0.48\textwidth]{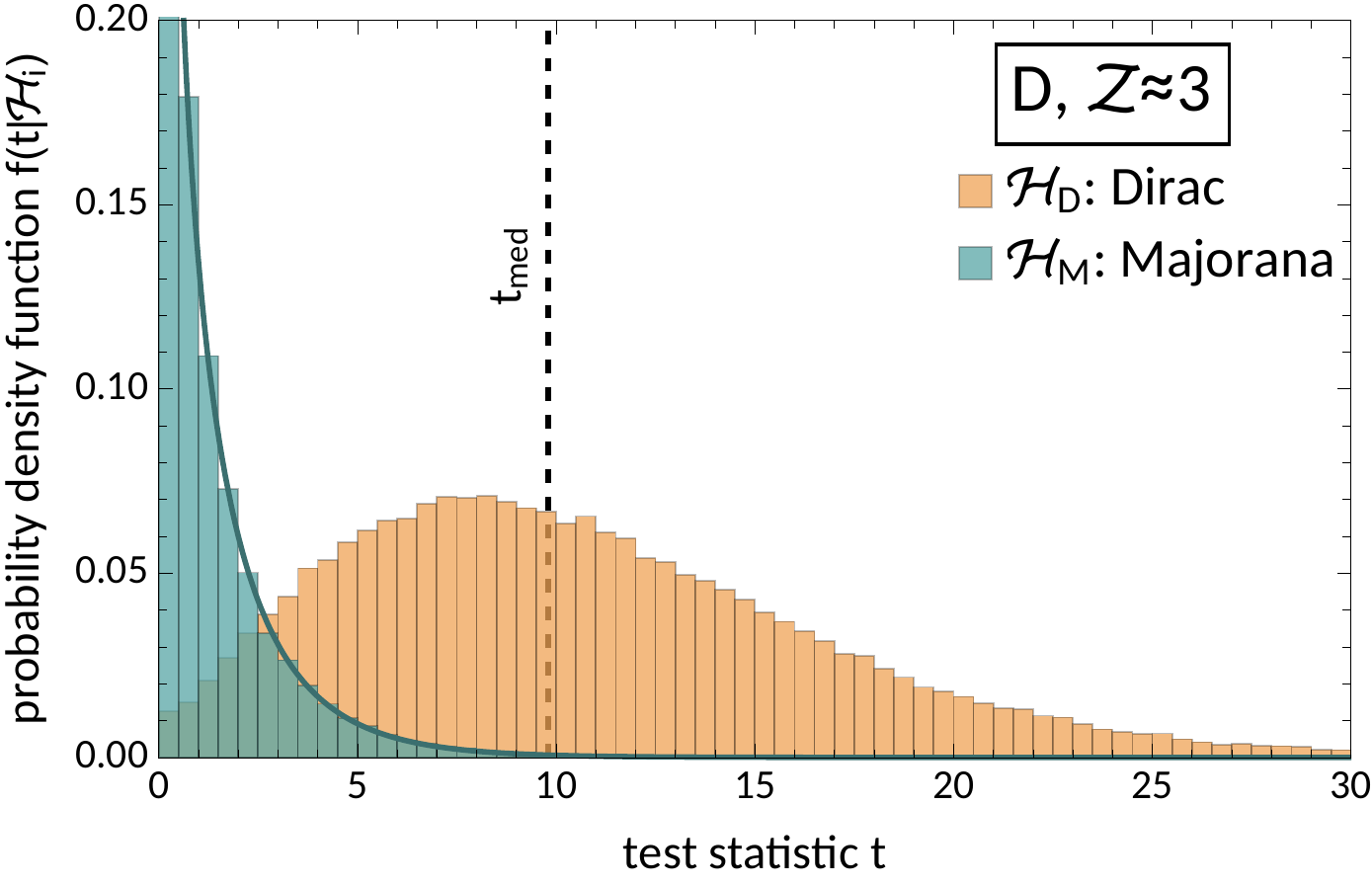}
\includegraphics[width=0.48\textwidth]{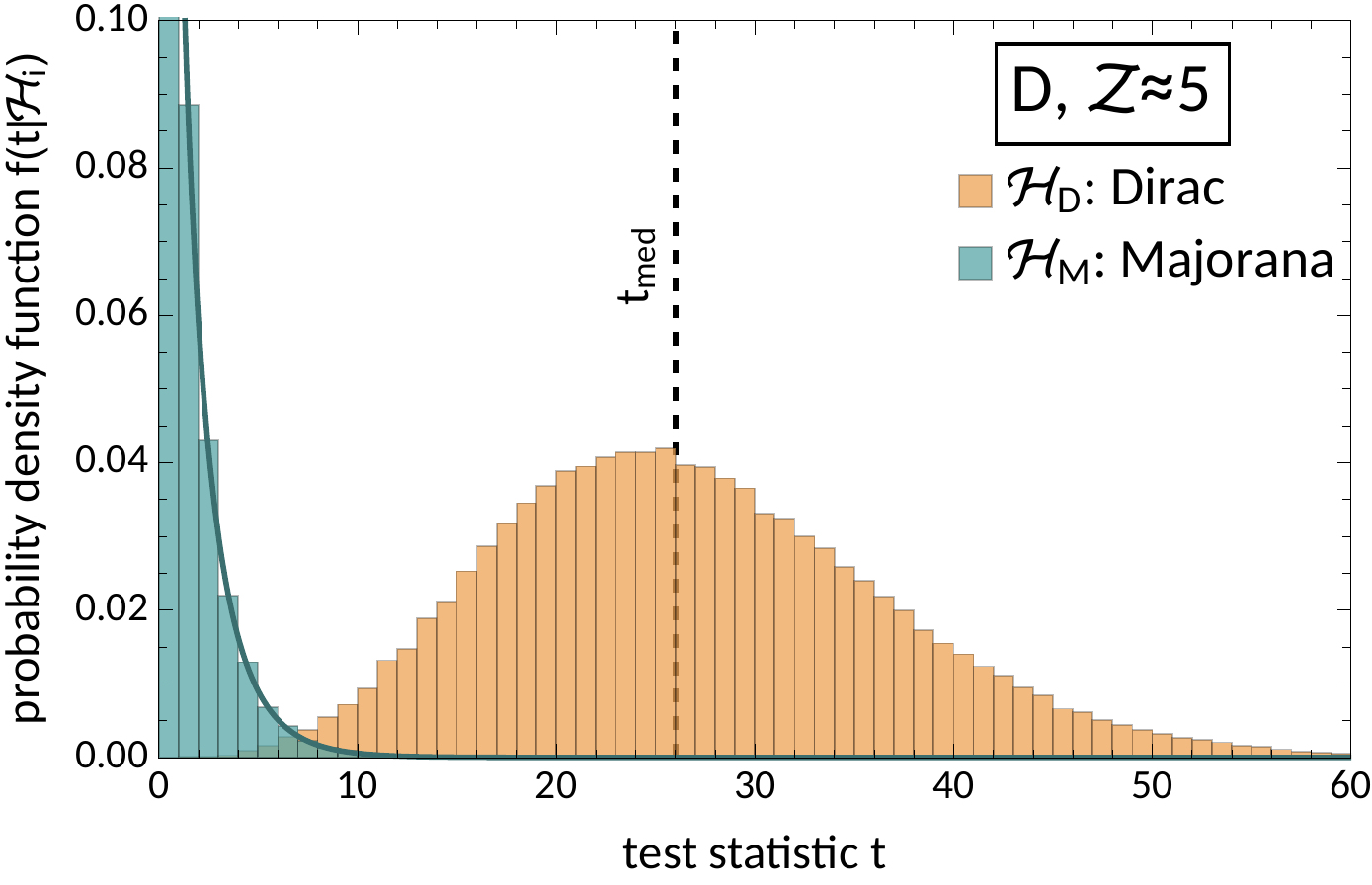}
\includegraphics[width=0.48\textwidth]{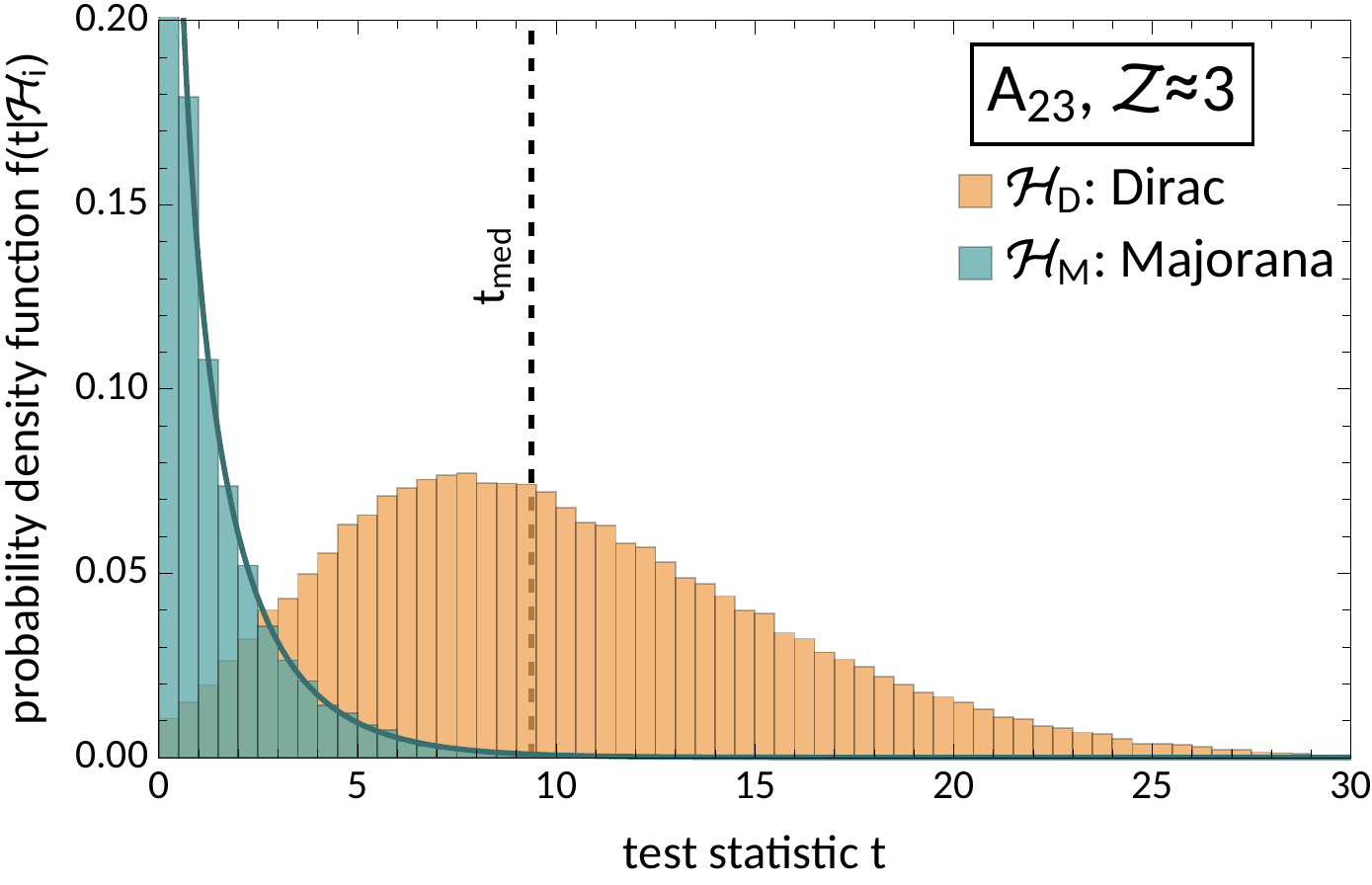}
\includegraphics[width=0.48\textwidth]{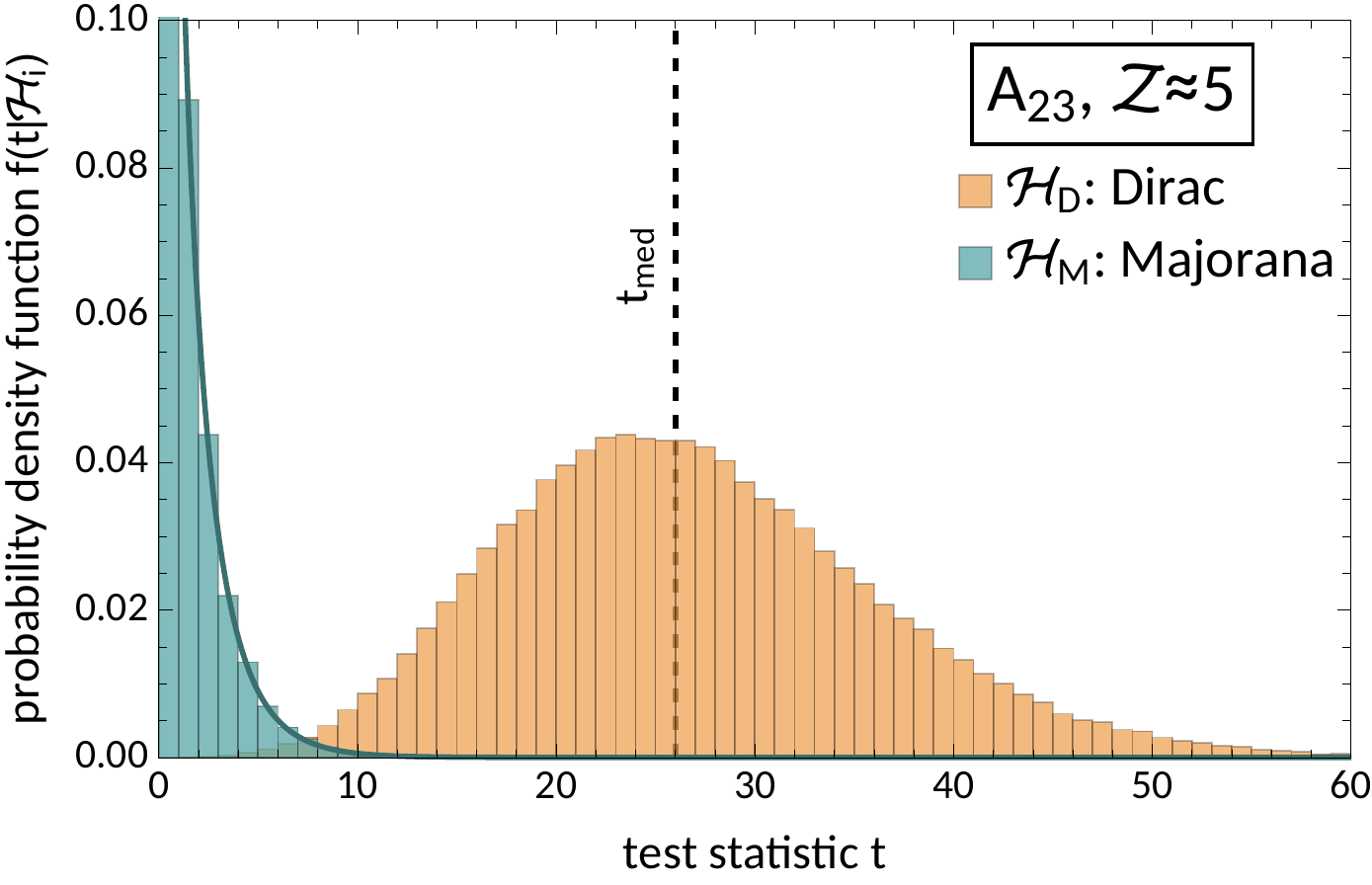}
\caption{Histogram estimate of the probability density function (pdf) $f(t|\mathscr{H}_i)$, $i=M, D$ of the test statistic~$t$ for the benchmark hierarchies D and $\text{A}_{23}$ (upper and lower row). The number of events has been chosen such that the expected significance to reject the null hypothesis is $\mathcal{Z}=3,5$ going from left to right.  The histograms show the MC based pdf (obtained by sampling the test statistic $10^5$ times). The continuous line shows the asymptotic pdf underlying Eq.~\eqref{eq:wi}.}
\label{fig: t distribution}
\end{figure}

The statistical procedure was introduced in detail in Sec.~\ref{sec:stat}. 
The probability density function of the test statistic~$t$, as defined in Eq.~\eqref{eq:q}, can be obtained by means of MC~simulations.
By simulating a great number of possible outcomes of the detection experiment in the form of signal data~$\mathscr{D}$, where one of the two hypotheses has to be assumed, and maximising the likelihoods of~$\mathscr{H}_M$ and~$\mathscr{H}_D$, we obtain a sample set of~$t$.
The resulting histogram estimates of~$f(t|\mathscr{H}_M)$ and~$f(t|\mathscr{H}_D)$ for two of the benchmark points (D and $\text{A}_{23}$) are depicted in Fig.~\ref{fig: t distribution}.
For this figure, the number of events was fixed to the value corresponding to a significance of~$\mathcal{Z}=3,5$, as we will discuss further below.

The first step is to determine the median~$t_{\rm med}$ of the distribution~$f(t|\mathscr{H}_D)$ under the alternative hypothesis, which is indicated as a vertical dashed line in Fig.~\ref{fig: t distribution}. As we also discussed in the previous section, there are two possible ways to obtain the $p$-value, or equivalently the statistical significance~$\mathcal{Z}$, based on the obtained histograms.

\begin{enumerate}
\item One possible way to solve the integral in Eq.~\eqref{eq:pvalue} is MC integration. The $p$-value corresponds to the relative amount of~$t$ values sampled under the null hypothesis which fall above~$t_{\rm med}$.
This fully MC-based procedure is the method of choice for lower numbers of signal events and thereby also higher p-values, where the exact distribution is generally unknown prior to MC simulations.
\item Using MC integration is practically inapplicable for the case of larger number of signal events and higher statistical significance, since the $t$~sample size under the null hypothesis necessary to have a reliable estimate of the $p$-value will be enormous.
Under these conditions, it is beneficial, and also appropriate, to use the asymptotic form of~$f(t|\mathscr{H}_M)$, which was discussed in detail in Sec.~\ref{sec:stat}.
In this case, $t$ obeys a~$\bar{\chi}^2$ distribution, consistently with Eq.~\eqref{eq:wi}.
This method can be regarded as a hybrid between MC and analytic methods, since the determination of the weights~$w_i$ in Eq.~\eqref{eq:wi} still require a MC sample of~$t$ under the null hypothesis, see Eqs.~\eqref{eq:system}.
Once the weights are known, $f(t|\mathscr{H}_M)$ can be integrated numerically to obtain the $p$-value.
\end{enumerate}

\begin{figure}
\centering
\includegraphics[width=\textwidth]{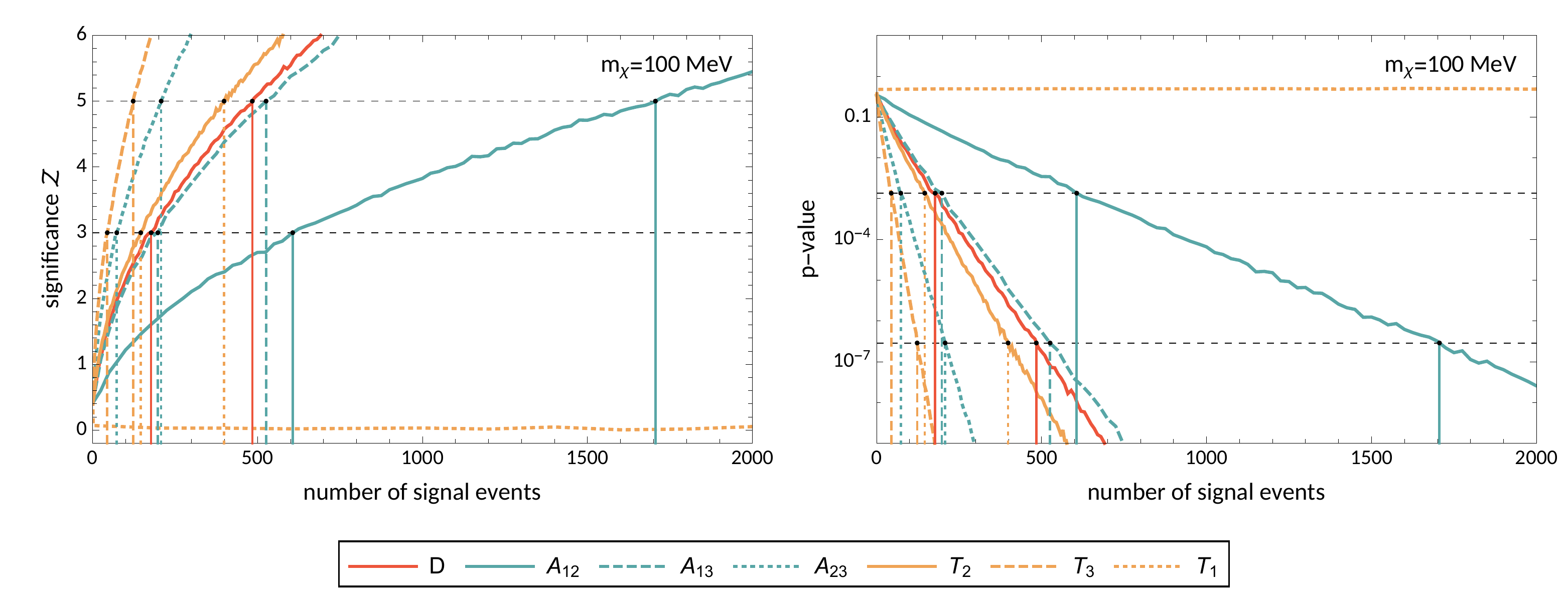}
\caption{Significance and p-values for the discrimination of Majorana and Dirac~DM as a function of number of observed signal events. The black dashed lines indicate the thresholds for the classification as \textit{evidence}($\mathcal{Z}=3$) or \textit{discovery}($\mathcal{Z}=5$). The vertical lines indicate the corresponding number of signals for the different benchmark hierarchies, which are also listed in Tab.~\ref{tab: required signals}.}
\label{fig: significance}
\end{figure}

Both procedures result in the expected Majorana DM rejection significance~$\mathcal{Z}$, defined in Eq.~\eqref{eq:Z}, as a function of observed signal events for the seven different Dirac scenarios.
This is shown in the left panel of Fig.~\ref{fig: significance}.
The right panel of the figure shows the equivalent evolution of the expected $p$-value as defined in Eq.~\eqref{eq:pvalue}.

The results confirm our qualitative expectations which were solely based on the spectral shapes in Fig.~\ref{fig: spectra}.
The Dirac hierarchy which resemble the null hypothesis to a larger degree by involving sizable contributions of anapole interactions, i.e. $\text{A}_{12}$ and $\text{A}_{13}$, are hardest to distinguish from the Majorana hypothesis (with the exception of $\text{T}_{1}$).
If DM was for example a Dirac fermion interacting dominantly via anapole and magnetic dipole interactions, it would require $\sim 610$ ($\sim 1700$) observed signals to be able to reject the Majorana nature with $\mathcal{Z}=3$ ($\mathcal{Z}=5$).

A scenario more favourable for us would be Dirac~DM with significant electric dipole interactions, which correspond to our benchmarks~$\text{A}_{23}$ and especially~$\text{T}_3$.
In the latter case, only $\sim 45$ ($\sim 120$) observed signals are expected to be sufficient to reject the scenario of Majorana~DM with statistical significance corresponding to 3(5) standard deviations.

\begin{figure}[h!]
\begin{floatrow}
\ffigbox{%
 \includegraphics[width=0.4\textwidth]{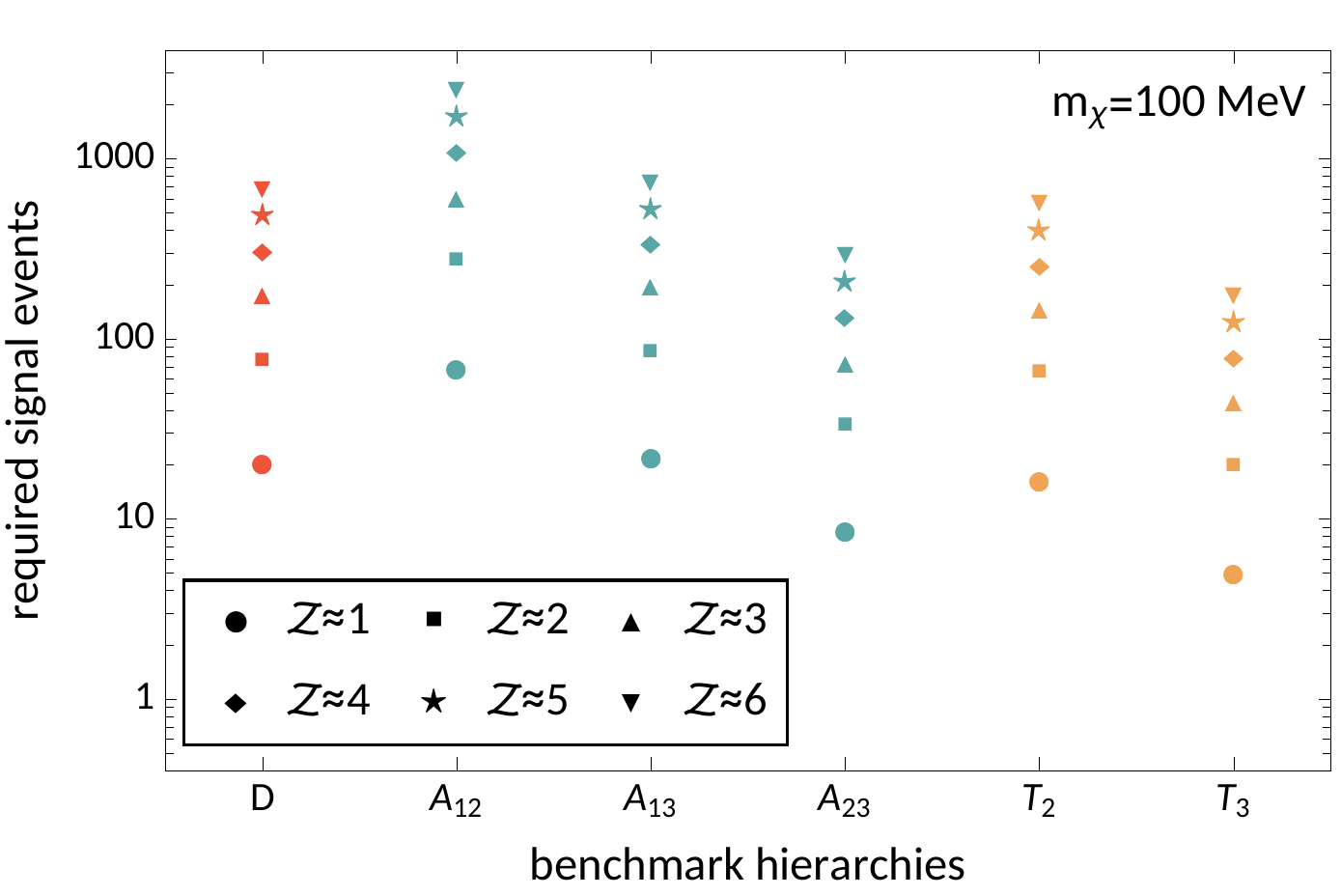}
}{%
  \caption{Number of signal events required to expect a statistical significance of $\mathcal{Z}$ for the rejection of the Majorana DM hypothesis.}%
  \label{fig: required events}
}
\capbtabbox{%
\begin{tabular}{|l|l|l|}
\hline
	&$\mathcal{Z}=3$	&$\mathcal{Z}=5$ \\
\hline
\hline
D 	&180	&480 \\
$\text{A}_{12}$ 	&610	&1700 \\
$\text{A}_{13}$ 	&200	&530 \\
$\text{A}_{23}$ 	&74	&210 \\
$\text{T}_{2}$ 	&150	&400 \\
$\text{T}_{3}$ 	&45	&120 \\
($\text{T}_{1}$) 	&($\gg 1000$)	&($\gg 1000$)\\ 
\hline
\end{tabular}
}{%
  \caption{Number of events required for an expected statistical significance of~$\mathcal{Z}=3(5)$ for the different hierarchies of the Dirac hypothesis (with $m_\chi=100$ MeV).}%
  \label{tab: required signals}
}
\end{floatrow}
\end{figure}

Finally, we sum up the complete results in Fig.~\ref{fig: required events} and Tab.~\ref{tab: required signals}, where we present the number of signals for each of our benchmark hierarchies necessary for an expected statistical significance~$\mathcal{Z}=1,...,6$ to reject the Majorana null hypothesis in favour of the Dirac hypothesis.

\section{Conclusions}
\label{sec:conclusions}

We have shown that if DM interacts with the Standard Model mainly via its higher-order electromagnetic moments and has a mass in the MeV-GeV range, direct detection experiments searching for atomic ionisation events induced by DM-electron scattering in xenon targets can shed some light on whether DM is its own antiparticle or not.~In support of this statement, we calculated the number of DM-induced atomic ionisations required to reject a scenario where DM is a Majorana particle in favour of an alternative scenario where DM has a Dirac nature, under the assumption the DM is a spin-1/2 fermion.~We found that the two scenarios can in principle be discriminated in case of DM discovery at direct detection experiments because the amplitude of the DM higher-order electromagnetic moments depends on whether DM is a Dirac or a Majorana particle.~More specifically, if DM is a Majorana particle, the anapole moment is its leading electromagnetic moment, since the magnetic dipole and the electric dipole are odd under particle-antiparticle exchange and therefore vanish.~In contrast, the electric dipole, the magnetic dipole and the electromagnetic anapole can simultaneously contribute to the DM-electron scattering in xenon detectors for Dirac DM.

Quantitatively, we found that between about 45 (130) and 610 (1700) DM signal events are required to reject the Majorana DM hypothesis in favour of the alternative Dirac hypothesis with a statistical significance corresponding to 3 (5) standard deviations.~The exact number of required DM signal events depends on the relative size of the anapole, magnetic dipole and electric dipole contributions to the expected rate of DM-induced atomic ionisations under the Dirac hypothesis.~For example, we found that statistically rejecting the Majorana hypothesis when the electric dipole moment is the leading coupling of Dirac DM (the $\text{T}_3$ scenario) is easier than in cases where either another coupling dominates, or different contributions to the signal event rate are comparable.~At the same time, rejecting the Majorana hypothesis in favour of a Dirac DM candidate coupling to the Standard Model via a linear combination of anapole and magnetic dipole interactions of similar strength (the $\text{A}_{12}$ scenario) appears to be more difficult as compared to other scenarios.~In this context, the worst-case-scenario is the one where DM is a Dirac particle and its leading coupling to the photon is the anapole moment~($\text{T}_1$).

Our results rely on Monte Carlo simulations and on the likelihood ratio as a test statistic.~While this method is standard in many applications, here it required a non-trivial extension to hypotheses that lie on the boundary of the parameter space. Indeed, this is the case for the Majorana DM hypothesis, which is characterised by two parameters of interest being zero.

This work could be further extended in various ways.~One possibility could be to include other target materials in the analysis, such as argon, for which the atomic responses generated by the anapole, magnetic and electric dipoles are already available~\cite{Catena:2019gfa}, or crystals, which are known to be sensitive to sub-GeV and even sub-MeV DM.~Furthermore, the analysis could be extended by including the DM particle mass as an additional nuisance parameter in our implementation of the likelihood ratio test.~At the same time, the results presented here constitute a solid proof of concept about the possibility of using direct detection experiments to reject the Majorana DM hypothesis when the DM interactions are dominated by higher-order electromagnetic moments.

\acknowledgments 
During this work, RC and TE were supported by the Knut and Alice Wallenberg Foundation (PI, Jan Conrad).~RC also acknowledges support from an individual research grant from the Swedish Research Council, dnr. 2018-05029.~The research presented in this article made use of the computer programme Wolfram Mathematica~\cite{Mathematica}.

\appendix

\section{Scattering amplitudes}
\label{sec:amplitudes}
Here, we calculate the amplitude for DM-electron scattering for the anapole, magnetic dipole and electric dipole interaction models, providing an explicit derivation for Eqs.~(\ref{eq:MNRa}), (\ref{eq:MNRm}) and (\ref{eq:MNRe}).~Conventions for the summation of repeated indexes are:~$L_\mu M^\mu =M^0L^0 - \mathbf{M} \cdot \mathbf{L}$, with $\mathbf{M} \cdot \mathbf{L} = M^\ell L^\ell = - g_{i j} M^i L^j = -M^i L_i$, where $g_{ij}=-\delta_{ij}$, $L^\mu = (L^0, \mathbf{L})$ and $M^\mu = (M^0, \mathbf{M})$ are arbitrary four-vectors, while $\mathbf{L}$ and $ \mathbf{M}$ are their three-dimensional space components.~We define the DM-electron scattering amplitude, $\mathcal{M}$, in terms of the corresponding $S$-matrix element,
\begin{align}
S & \equiv \frac{1}{\sqrt{2 E_{\mathbf{p}}2 E_{\mathbf{p}'}2 E_{\mathbf{k}}2 E_{\mathbf{k}'} }}  (2 \pi)^4 \delta^{(4)}(p'+k'-p-k)\, i\mathcal{M} \nonumber\\
&= i \int {\rm d}^4x \langle \mathbf{p}',s',\mathbf{k}',r'| \mathscr{L}_I(x)|\mathbf{p},s,\mathbf{k},r \rangle
\label{eq:M}
\end{align}
where $\mathscr{L}_I(x)$ is the assumed interaction Lagrangian.~Here, $p$ denotes the four-momentum associated with the tridimensional momentum $\mathbf{p}$, i.e.~$p^\mu=(E_{\mathbf{p}},\mathbf{p})$, where $E_{\mathbf{p}}\equiv \sqrt{|\mathbf{p}|^2+m_\chi^2}$,  and similarly for $k$, $p'$ and $k'$.~Initial and final state are defined as tensor products of single-particle states, e.g.~$|\mathbf{p},s,\mathbf{k},r \rangle = |\mathbf{p},s\rangle \otimes |\mathbf{k},r\rangle$.~These are labelled by the initial (final) state DM and electron momenta, $\mathbf{p}$ and $\mathbf{k}$ ($\mathbf{p}'$ and $\mathbf{k}'$) and by the initial (final) state DM and electron spins, $s$ and $r$ ($s'$ and $r'$), respectively.~Following~\cite{Catena:2019gfa}, single-particle states are normalised as follows 
\begin{align}
\langle \mathbf{p}',s' | \mathbf{p},s\rangle = \delta_{s's} \,\delta^{3}(\mathbf{p}'-\mathbf{p})\,.
\label{eq:norm}
\end{align}
For each of the three models considered here, the starting point for our calculation is to express the interaction Lagrangian in terms of a local DM current, $j^{\mu}(x)$, and the photon field, $A_\mu(x)$,
\begin{align}
\mathscr{L}_I(x) = -j^{\mu}(x) A_{\mu}(x)\,,
\label{eq:Lcurrent}
\end{align}
by means of partial integration.~By doing so, the $S$-matrix element in Eq.~(\ref{eq:M}) can be written as follows
\begin{align}
S &= -i \int {\rm d}^4x \langle \mathbf{p}',s'| j^{\mu}(x)|\mathbf{p},s \rangle \langle \mathbf{k}',r'|A_{\mu}(x) | \mathbf{k},r \rangle\nonumber\\
&= -i\langle \mathbf{p}',s'| j^{\mu}(0)|\mathbf{p},s \rangle \langle \mathbf{k}',r'|A_{\mu}(q) | \mathbf{k},r \rangle \,,
\label{eq:M2}
\end{align}
where in the second step we translated the DM current from the spacetime point $x$ to the origin, $j^{\mu}(x)=e^{i \hat{P}\cdot x} j^{\mu}(0) e^{-i \hat{P}\cdot x}$, $\hat{P}^\mu$ being the four-dimensional momentum operator, and introduced the Fourier transform of the photon field, $A_\mu(q) = \int {\rm d}^4x e^{-i q \cdot x} A_{\mu} (x)$.~Notice that $|\mathbf{p},s \rangle$ ($|\mathbf{p}',s' \rangle$) is an eigenstate of energy and momentum and therefore $\hat{P}^\mu|\mathbf{p},s \rangle = p^\mu |\mathbf{p},s \rangle$ ($\hat{P}^\mu|\mathbf{p}',s' \rangle = p^{' \mu} |\mathbf{p}',s' \rangle$).~Here, we also introduced the momentum transfer, defined as $q=p-p'$.~The matrix element $ \langle \mathbf{k}',r'|A_{\mu}(q) | \mathbf{k},r \rangle$ is model independent, i.e. it only depends on QED, and can therefore be expressed in terms of the electromagnetic current, $J_\mu(x)$, at the origin,
\begin{align}
\langle \mathbf{k}',r'|A_{\mu}(q) | \mathbf{k},r \rangle = - (2 \pi)^4 \delta^{(4)}(p'+k'-p-k)\, \frac{ e \langle \mathbf{k}',r'|J_{\mu}(0) | \mathbf{k},r \rangle}{q^2} \,.
\label{eq:A2J}
\end{align}
Here, we used Maxwell equations, $\partial_\mu F^{\mu \nu} = e J^{\nu}$, the definition of $F_{\mu \nu}=\partial_\mu A_\nu - \partial_\nu A_\mu$, and the identity
\begin{align}
\partial_\mu F^{\mu \nu} (x)= - \int {\rm d}^4 q \,e^{i q \cdot x} \left[ q^2 A^\nu (q) -q^\nu q_\mu A^\mu(q) \right] \,,
\end{align}
with $q_\mu A^\mu = 0$.~The four-dimensional Dirac delta in Eq.~(\ref{eq:A2J}) arises from translating $J_\mu(x)$ to the origin,
\begin{align}
\int {\rm d}^4 x\, e^{-i q\cdot x} \langle \mathbf{k}',r'|J_{\mu}(x) | \mathbf{k},r \rangle &=  \int {\rm d}^4 x\, e^{-i q\cdot x} \langle \mathbf{k}',r'|e^{i \hat{P}\cdot x} J_{\mu}(0) e^{-i \hat{P}\cdot x} | \mathbf{k},r \rangle \nonumber\\
&=\int {\rm d}^4 x\, e^{-i (p-p'+k-k')\cdot x} \langle \mathbf{k}',r'|J_{\mu}(0) | \mathbf{k},r \rangle \nonumber\\
&=(2 \pi)^4 \delta^{(4)}(p'+k'-p-k) \langle \mathbf{k}',r'|J_{\mu}(0) | \mathbf{k},r \rangle\,.
\end{align}
Notice that the matrix element $\langle \mathbf{k}',r'|J_{\mu}(0) | \mathbf{k},r \rangle$ can be expressed in terms of electromagnetic form factors,
\begin{align}
\langle \mathbf{k}',r'|J_{\mu}(0) | \mathbf{k},r \rangle = \frac{1}{\sqrt{2 E_{\mathbf{k}} 2 E_{\mathbf{k}'}}} \bar{v}^{r'}(k') \left[ F_1(q^2) \gamma_\mu + \frac{i}{2 m_e} F_2(q^2) \sigma_{\mu \nu} q^\nu \right] v^r(k) \,.
\label{eq:matrixJ}
\end{align}
Here, $F_1(0)=1$, $g_e=2[F_1(0)+F_2(0)]\simeq 2$ is the electron $g$-factor, and the four-component spinor $v^r(\mathbf{k})$ is a solution to the free field Dirac equation for the electron.~In order to extract the amplitude $\mathcal{M}$ from Eq.~(\ref{eq:M2}), the next step is to evaluate the matrix element $\langle \mathbf{p}',s'| j^{\mu}(0)|\mathbf{p},s \rangle$.~This part of the calculation is model dependent and will be presented below for the anapole (Sec.~\ref{sec:a}) magnetic dipole (Sec.~\ref{sec:m}) and electric dipole (Sec.~\ref{sec:e}) DM couplings separately.

\subsection{Anapole}
Let us start from the case of Majorana DM with anapole interactions, and focus on Dirac DM subsequently.~The interaction Lagrangian for this model is given by Eq.~(\ref{eq:LIa}).~This Lagrangian is equivalent to Eq.~(\ref{eq:Lcurrent}) plus a total derivative if we define the local interaction current $j^\mu(x)$ as follows
\label{sec:a}
\begin{align}
j^{\mu}(x) = -\frac{g_{1}}{2 \Lambda^2} \left( g^{\mu \lambda} \partial^\nu \partial_\nu - \partial^\mu \partial^\lambda \right) \bar{\chi}(x) \gamma_\lambda \gamma_5 \chi(x)\,.
\label{eq:caM}
\end{align}
The Majorana spinor field $\chi(x)$ can be expanded in terms of a single set of creation and annihilation operators, $a_{\mathbf{p}}^s$ and $a_{\mathbf{p}}^{s \dagger}$, respectively, and a four-component spinor, $u^{s}(p)$, solving the free field Dirac equation with $\gamma$-matrices in the Majorana representation ($\gamma^{\mu *} = - \gamma^{\mu}$ and $\gamma_5^* = -\gamma_5$),
\begin{align}
\chi(x) = \sum_s \int \frac{{\rm d}^3 p}{(2 \pi)^3} \frac{1}{\sqrt{2 E_{\mathbf{p}}}} \bigg( e^{-i p\cdot x} u^s(p) a_{\mathbf{p}}^s + e^{i p\cdot x} u^{s *}(p) a_{\mathbf{p}}^{s \dagger}\bigg) \equiv \chi^{+}(x) + \chi^{-}(x)\,.
\end{align}
In this expansion, we denote by $\chi^{(+)}(x)$ the term proportional to $e^{-i p\cdot x}$ and by $\chi^{(-)}(x)$ the term proportional to $e^{+i p\cdot x}$.~Similarly, we decompose $\bar{\chi}=\chi^\dagger \gamma^0$ as $\bar{\chi}(x)=\bar{\chi}^{(+)}(x)+\bar{\chi}^{(-)}(x)$, where $(+)$ and $(-)$ refer to positive and negative frequency solutions of the Dirac equation, as usual in second quantisation.~With this notation, $\langle \mathbf{p}',s'| j^{\mu}(0)|\mathbf{p},s \rangle $ can be expressed as follows
\begin{align}
\langle \mathbf{p}',s'| j^{\mu}(0)|\mathbf{p},s \rangle &=  -\frac{g_{1}}{2 \Lambda^2} \left( g^{\mu \lambda} \partial^\nu \partial_\nu - \partial^\mu \partial^\lambda \right) \langle 0| a_{\mathbf{p}'}^{s' }\bigg[ \bar{\chi}^{(-)}(x) \gamma_\lambda \gamma_5 \chi^{(+)}(x) \nonumber \\
&+ \bar{\chi}^{(+)}(x) \gamma_\lambda \gamma_5 \chi^{(-)}(x)\bigg] a_{\mathbf{p}}^{s\dagger} | 0 \rangle \,,
\label{eq:2contributions}
\end{align}
where $|\mathbf{p},s\rangle=a_{\mathbf{p}}^{s\dagger} | 0 \rangle$, $\langle \mathbf{p}',s'|=\langle 0|a_{\mathbf{p}'}^{s'}$ and $\{a_{\mathbf{p}}^{s},a_{\mathbf{p}'}^{s'\dagger}\}=(2\pi)^3 \delta_{s s'} \delta^{(3)}(\mathbf{p}-\mathbf{p}')$, consistently with Eq.~(\ref{eq:norm}).~The two terms in Eq.~(\ref{eq:2contributions}) give identical contributions to the matrix element.~We find
\begin{align}
\langle \mathbf{p}',s'| j^{\mu}(0)|\mathbf{p},s \rangle &= \frac{1}{\sqrt{2 E_{\mathbf{p}}2 E_{\mathbf{p}'} }}\frac{g_1}{\Lambda^2} q^2 \left( g^{\mu\lambda} -\frac{q^\mu q^\lambda}{q^2}  \right)
\, \bar{u}^{s'}(p')\gamma_\lambda \gamma_5 u^s(p) \,,
\label{eq:matrixj}
\end{align}
where we used $\left[u^{s'T}(p')\gamma^0\gamma_\lambda \gamma_5 u^{s *}(p)\right]^*=-\bar{u}^{s'}(p')\gamma_\lambda \gamma_5 u^s(p)$ ($u^{s'T}(p')$ is the transpose of $u^{s'}(p')$). Combining Eqs.~(\ref{eq:matrixJ}) and (\ref{eq:matrixj}) with Eq.~(\ref{eq:M}), for Majorana DM with anapole interactions we obtain
\begin{align}
\mathcal{M} &= \frac{eg_1}{\Lambda^2} \left( g^{\mu\lambda} -\frac{q^\mu q^\lambda}{q^2}  \right) 
 \bar{u}^{s'}(p')\gamma_\lambda \gamma_5 u^s(p) \,\bar{v}^{r'}(k') \left[ F_1(q^2) \gamma_\mu 
+ \frac{i}{2 m_e} F_2(q^2) \sigma_{\mu \nu} q^\nu \right] v^r(k) \,.
\label{eq:Ma}
\end{align}
In the non-relativistic limit, 
\begin{align}
\mathcal{M} &\simeq \frac{eg_1}{\Lambda^2} \bigg[ F_1(0) g^{\mu\lambda}  \,\, \bar{u}^{s'}(p')\gamma_\lambda \gamma_5 u^s(p)  \,\,\bar{v}^{r'}(k')\gamma_\mu v^r(k) 
\nonumber\\
& + \frac{i}{2 m_e} F_2(0) g^{ij} \bar{u}^{s'}(p')\gamma_j \gamma_5 u^s(p) \bar{v}^{r'}(k') \sigma_{i \nu } q^\nu v^r(k) \bigg] \nonumber\\
&\simeq \frac{eg_1}{\Lambda^2}  \bigg\{ \bigg[ 2(\mathbf{p}+\mathbf{p}') \cdot \xi^{s'} \frac{\boldsymbol{\sigma}}{2} \xi^s \bigg] 2 m_e F_1(0) \delta^{r' r} \nonumber \\
&-4 m_\chi F_1(0) \xi^{s'\dagger} \frac{\boldsymbol{\sigma}}{2}\xi^s \cdot \bigg[ (\mathbf{k}+\mathbf{k}')\delta^{r' r} - 2i \mathbf{q} \times \eta^{r'} \frac{\boldsymbol{\sigma}}{2} \eta^r  \bigg] \nonumber\\
&+4 m_\chi F_2(0) \xi^{s'\dagger} \frac{\boldsymbol{\sigma}}{2}\xi^s \cdot \left( 2i \mathbf{q} \times \eta^{r'} \frac{\boldsymbol{\sigma}}{2} \eta^r  \right) \bigg\}\,,
\label{eq:}
\end{align}
where we expanded the spinors $u^s(p)\simeq ((2m_\chi - \mathbf{p}\cdot \boldsymbol{\sigma})\xi^s, (2m_\chi + \mathbf{p}\cdot \boldsymbol{\sigma})\xi^s)^T/\sqrt{4m_\chi}$ and $v^r(k)\simeq ((2m_\chi - \mathbf{k}\cdot \boldsymbol{\sigma})\eta^r, (2m_\chi + \mathbf{k}\cdot \boldsymbol{\sigma})\eta^r)^T/\sqrt{4m_e}$ and their spinor bilinears in the non-relativistic limit, 
\begin{align}
\bar{u}^{s'}(p')\gamma_\lambda \gamma_5 u^s(p) &\simeq \left( 2(\mathbf{p}+\mathbf{p}') \cdot \xi^{s'} \frac{\boldsymbol{\sigma}}{2} \xi^s , -4 m_\chi \, \xi^{s'} \frac{\boldsymbol{\sigma}}{2} \xi^s  \right)^T \nonumber\\
\bar{v}^{r'}(k')\gamma_\mu v^r(k) &\simeq  \left( 2m_e \delta^{r' r}, - (\mathbf{k}+\mathbf{k}')\delta^{r' r}  + 2i \mathbf{q} \times \eta^{r'} \frac{\boldsymbol{\sigma}}{2} \eta^r  \right)^T \nonumber\\
\bar{v}^{r'}(k') g^{ij}\sigma_{i \nu } q^\nu v^r(k)  &\simeq - q^\ell \bar{v}^{r'}(k') \sigma^{j \ell } v^r(k) \simeq -4 m_e q^\ell \epsilon^{j\ell k} \eta^{r'} \frac{\sigma^k}{2} \eta^r \,.
\label{eq:bilinear1}
\end{align}
Notice that the bilinear spinor expansions in Eq.~(\ref{eq:bilinear1}) apply to all $\gamma$-matrix and spinor representations that are related to the Dirac representation by a unitary transformation.~For $\mathcal{M}$, we finally obtain
\begin{align}
\mathcal{M}&= \frac{4 e g_1}{\Lambda^2} m_\chi m_e \Bigg\{
    2 \left(\mathbf{v}_{\rm el}^\perp \cdot \xi^{ s'\dagger} \mathbf{S}_\chi \xi^s \right) \delta^{r' r} +
    g_e \left( \xi^{s'\dagger } \mathbf{S}_\chi \xi^s \right) \cdot \left( i\frac{\mathbf{q}}{m_e} \times \eta^{r'\dagger} \mathbf{S}_e \eta^r  \right) 
    \Bigg\}\,,
\end{align}
where both spin operators, $\mathbf{S}_\chi=\boldsymbol{\sigma}/2$ and $\mathbf{S}_e=\boldsymbol{\sigma}/2$, are defined in terms of the three Pauli matrices, $\boldsymbol{\sigma}=(\sigma^1, \sigma^2, \sigma^3)$.~In the former case, the Pauli matrices act on the DM particle spin space spanned by the two-component spinors $\xi^s$, $s=1,2$.~In the latter one, they act on the electron spin space, which is generated by the independent, two-component spinors $\eta^r$, $r=1,2$.

Let us now calculate the amplitude for DM-electron scattering for the case of Dirac DM coupling to photons via an anapole interaction.~In this case, the current $j^\mu(x)$  in Eq.~(\ref{eq:Lcurrent}) reads
\begin{align}
j^{\mu}(x) = -\frac{g_{1}}{\Lambda^2} \left( g^{\mu \lambda} \partial^\nu \partial_\nu - \partial^\mu \partial^\lambda \right) \bar{\psi}(x) \gamma_\lambda \gamma_5 \psi(x)\,,
\label{eq:caD}
\end{align}
where $\psi(x)$ is a Dirac spinor field.~As already mentioned, we assume that $g_1$ and $\Lambda$ are the same as in Eq.~(\ref{eq:caM}).~The overall factor of 2 difference between Eqs.~(\ref{eq:caM}) and (\ref{eq:caD}) arises from the Lagrangian of the two models.~In terms of creation and annihilation operators, $\psi(x)$ is given by
\begin{align}
\psi(x) = \sum_s \int \frac{{\rm d}^3 p}{(2 \pi)^3} \frac{1}{\sqrt{2 E_{\mathbf{p}}}} \bigg( e^{-i p\cdot x} u^s(p) a_{\mathbf{p}}^s + e^{i p\cdot x} u^{s *}(p) b_{\mathbf{p}}^{s \dagger}\bigg) \equiv \psi^{+}(x) + \psi^{-}(x)\,,
\label{eq:psi}
\end{align}
where $a_{\mathbf{p}}^s$ and $a_{\mathbf{p}}^{s\dagger}$, and $b_{\mathbf{p}}^s$ and $b_{\mathbf{p}}^{s\dagger}$ are independent sets of creation and annihilation operators.~The calculation of $\mathcal{M}$ in this case follows closely the one we made for Majorana DM with anapole interactions.~The only difference is that the second term in Eq.~(\ref{eq:2contributions}), $\langle 0 |a_{\mathbf{p}}^{s} \bar{\psi}^{(+)}(x) \gamma_\lambda \gamma_5 \psi^{(-)}(x) a_{\mathbf{p}}^{s\dagger} | 0 \rangle$, vanishes in the case of Dirac DM, as one can show by using Eq.~(\ref{eq:psi}).~This compensates for the factor of 2 difference in the currents of Dirac and Majorana DM, in that the two terms in Eq.~(\ref{eq:2contributions}) give identical contributions to the amplitude $\mathcal{M}$.~Summarising, for a given value of $g_1/\Lambda^2$ the Dirac and Majorana DM models with anapole interactions defined in Eqs.~(\ref{eq:LIa}) and ~(\ref{eq:LIm}), respectively, predict the same $\mathcal{M}$ and are indistinguishable.

\subsection{Magnetic dipole}
\label{sec:m}
In the case of Majorana DM, the interaction operator $\bar{\chi}\sigma^{\mu\nu}\chi F_{\mu \nu}$ vanishes exactly.~Only Dirac DM can couple to the photon via a magnetic dipole.~In this case, the current $j^\nu(x)$  in Eq.~(\ref{eq:Lcurrent}) reads
\begin{align}
j^\nu(x)= - \frac{2g_2}{\Lambda} \partial_{\mu} \left[ \,\bar{\psi}(x)\sigma^{\mu\nu} \psi(x) \right]\,,
\end{align}
where $\psi(x)$ is a Dirac spinor field with expansion in annihilation a creation operators given in Eq.~(\ref{eq:psi}).~For the matrix element of the current $j^\nu(x)$ between the states $| a^{s \dagger}_{\mathbf{p}}|0 \rangle$ and $a^{s' \dagger}_{\mathbf{p}'}|0 \rangle$, we find
\begin{align}
\langle \mathbf{p}',s'| j^{\nu}(0)|\mathbf{p},s \rangle &= -\frac{2g_{2}}{\Lambda} \partial_\mu \langle 0| a_{\mathbf{p}'}^{s' }\left[ \bar{\psi}^{(-)}(x) \sigma^{\mu \nu} \psi^{(+)}(x) \right] a_{\mathbf{p}}^{s\dagger} | 0 \rangle \big|_{x=0} \nonumber\\
&= \frac{2g_2}{\Lambda} \frac{1}{\sqrt{2 E_{\mathbf{p}} 2E_{\mathbf{p}'}}} \,iq_\mu \,\bar{u}^{s'}(p') \sigma^{\mu \nu} u^s(p) \,.
\label{eq:matrixjDM}
\end{align}
Substituting this expression in Eq.~(\ref{eq:M2}) and using Eq.~(\ref{eq:matrixJ}) for the matrix element of $J_\mu(x)$, we obtain 
\begin{align}
\mathcal{M}=-2ie \frac{g_2}{\Lambda} \frac{1}{|\mathbf{q}|^2} q_\mu \bar{u}^{s'}(p') \sigma^{\mu \nu} u^s(p) \,\bar{v}^{r'}(k') \left[ F_1(q^2) \gamma_\nu + \frac{i}{2 m_e} F_2(q^2) \sigma_{\nu \alpha} q^\alpha \right] v^r(k) \,,
\label{eq:Mm}
\end{align}
with $q^2=-|\mathbf{q}|^2$.~In the non-relativistic limit the transition amplitude in Eq.~(\ref{eq:Mm}) reads as follows
\begin{align}
\mathcal{M}&\simeq -2ie \frac{g_2}{\Lambda} \frac{1}{|\mathbf{q}|^2} q_i \bigg\{ F_1(0) \, \bar{u}^{s'}(p') \sigma^{i 0} u^s(p) \,\bar{v}^{r'}(k')  \gamma_0 v^r(k) \nonumber \\ &+\bar{u}^{s'}(p') \sigma^{i j} u^s(p) \,\bar{v}^{r'}(k') \bigg[ F_1(0) \gamma_j 
+ \frac{i}{2 m_e} F_2(0) \sigma_{j \ell} q^\ell \bigg] v^r(k) \bigg\} \nonumber\\
&\simeq -2ie \frac{g_2}{\Lambda} \frac{1}{|\mathbf{q}|^2} q_i \bigg\{ F_1(0)\bigg[ -i q^i \delta^{s' s} + \epsilon^{i \ell m} (\mathbf{p}+\mathbf{p}')^\ell \xi^{s'}\sigma^m\xi^s \bigg] 2 m_e \delta^{r' r} \nonumber\\
&+ g_{jk} \left(2m_\chi \epsilon^{ijm} \xi^{s'} \sigma^m \xi^s \right) \bigg[ F_1(0) \left( (\mathbf{k}+\mathbf{k}')^k\delta^{r' r} -i\epsilon^{k \ell m} q^\ell \eta^{r'}\sigma^m \eta^r \right) \nonumber\\
&- \frac{i}{2m_e} F_2(0) \left( 2 m_e \epsilon^{k \ell m} q^\ell \eta^{r'}\sigma^m \eta^r \right) \bigg] \bigg\} \,,
\label{eq:MmNR}
\end{align}
where, in addition to Eq.~(\ref{eq:bilinear1}), we used the following non-relativistic expansion for the spinor bilinear
\begin{align}
\bar{u}^{s'}(p') \sigma^{i 0} u^s(p) \simeq  -i q^i \delta^{s' s} + \epsilon^{i \ell m} (\mathbf{p}+\mathbf{p}')^\ell \xi^{s'}\sigma^m\xi^s\,.
\label{eq:bilinear2}
\end{align}
Manipulating Eq.~(\ref{eq:MmNR}) by using standard vectorial identities, i.e. $(\mathbf{a}\times\mathbf{b})\times\mathbf{c} = -(\mathbf{c}\cdot \mathbf{b}) \mathbf{a} + (\mathbf{c}\cdot \mathbf{a}) \mathbf{b}$, where $\mathbf{a}$, $\mathbf{b}$ and $\mathbf{c}$ are tridimensional vectors, we arrive at our final result for the non-relativistic limit of the DM-electron scattering amplitude for Dirac DM with magnetic dipole interactions,
\begin{align}
\mathcal{M}&=\frac{e g_2}{\Lambda}  \Bigg\{
    4m_e\delta^{s's}\delta^{r'r} +\frac{16m_\chi m_e}{|\mathbf{q}|^2}  i\mathbf{q} \cdot \left(\mathbf{v}_{\rm el}^\perp \times \xi^{s' \dagger} \mathbf{S}_\chi \xi^s \right)\delta^{r'r} 
    \nonumber\\
    &-  \frac{8 g_em_\chi}{|\mathbf{q}|^2} \Bigg[\left( \mathbf{q} \cdot \xi^{s'\dagger} \mathbf{S}_\chi \xi^s \right)\left( \mathbf{q} \cdot \eta^{r'\dagger } \mathbf{S}_e \eta^r \right)
    - |\mathbf{q}|^2  \left( \xi^{s'\dagger} \mathbf{S}_\chi \xi^s \right)\cdot \left( \eta^{r'\dagger } \mathbf{S}_e \eta^r \right)
    \Bigg] \Bigg\}\,.
\end{align}
\subsection{Electric dipole}
\label{sec:e}
Similarly to the case of magnetic dipole interactions, for Majorana DM the interaction operator $i\bar{\chi}\sigma^{\mu \nu} \gamma^5 \chi F_{\mu\nu}$ (the electric dipole) vanishes exactly.~In contrast, an electric dipole interaction between Dirac DM and photons is allowed.~In this case, the current $j^\nu(x)$  in Eq.~(\ref{eq:Lcurrent}) reads as follows
\begin{align}
j^\nu(x) = -\frac{2g_3}{\Lambda} \,i\partial_{\mu} \left[\,\bar{\psi}(x) \sigma^{\mu \nu} \gamma^5 \psi(x)\right] \,.
\end{align}
Analogously to Eq.~(\ref{eq:matrixjDM}), for the matrix element of $j^\nu(x)$ between the states $| a^{s \dagger}_{\mathbf{p}}|0 \rangle$ and $a^{s' \dagger}_{\mathbf{p}'}|0 \rangle$, we find
\begin{align}
\langle \mathbf{p}',s'| j^{\nu}(0)|\mathbf{p},s \rangle &= -\frac{2g_{3}}{\Lambda}\,i \partial_\mu \langle 0| a_{\mathbf{p}'}^{s' }\left[ \bar{\psi}^{(-)}(x) \sigma^{\mu \nu} \gamma^5 \psi^{(+)}(x) \right] a_{\mathbf{p}}^{s\dagger} | 0 \rangle \big|_{x=0} \nonumber\\
&= - \frac{2g_3}{\Lambda} \frac{1}{\sqrt{2 E_{\mathbf{p}} 2E_{\mathbf{p}'}}} q_\mu \,\bar{u}^{s'}(p') \sigma^{\mu \nu} \gamma^5u^s(p) \,.
\end{align}
Combining the above expression with Eq.~(\ref{eq:matrixJ}), for Dirac DM with electric dipole interactions we obtain
\begin{align}
\mathcal{M}=2e \frac{g_3}{\Lambda} \frac{1}{|\mathbf{q}|^2} q_\mu \bar{u}^{s'}(p') \sigma^{\mu \nu} \gamma^5 u^s(p) \,\bar{v}^{r'}(k') \left[ F_1(q^2) \gamma_\nu + \frac{i}{2 m_e} F_2(q^2) \sigma_{\nu \alpha} q^\alpha \right] v^r(k) \,,
\end{align}
which in the non-relativistic limit reduces to
\begin{align}
\mathcal{M} &\simeq e \frac{2g_3}{\Lambda} \frac{1}{|\mathbf{q}|^2} \, q_i \left[ \bar{u}^{s'}(p') \sigma^{i 0} \gamma^5 u^s(p) \right] \left[  \bar{v}^{r'}(k') F_1(0) \gamma_0  v^r(k) \right] \nonumber\\
&\simeq e \frac{2g_3}{\Lambda} \frac{1}{|\mathbf{q}|^2} \, q_i  \left( -i 4 m_\chi\, \xi^{s'\dagger}\frac{\sigma^i}{2}\xi^s \right) 2 m_e F_1(0) \delta^{r' r}  \,,
\label{eq:Mtmp}
\end{align}
where we used the second expression in Eq.~(\ref{eq:bilinear1}), Eq.~(\ref{eq:bilinear2}), and the non-relativistic expansions
\begin{align}
\bar{u}^{s'}(p') \sigma^{i 0} \gamma^5 u^s(p) &\simeq -i 4 m_\chi\, \xi^{s'\dagger}\frac{\sigma^i}{2}\xi^s \nonumber\\
\bar{u}^{s'}(p') \sigma^{i j} \gamma^5 u^s(p) &\simeq \epsilon^{ijk}q^k + 2i (\mathbf{p} + \mathbf{p}')^i  \xi^{s'} \frac{\sigma^j}{2}\xi^s
- 2i (\mathbf{p} + \mathbf{p}')^j  \xi^{s'} \frac{\sigma^i}{2}\xi^s\,.
\end{align}
Eq.~(\ref{eq:Mtmp}) leads to our final result for the amplitude for DM-electron scattering for electric dipole Dirac DM,
\begin{align}
\mathcal{M}&= \frac{e g_3}{\Lambda} \frac{16 m_\chi m_e}{|\mathbf{q}|^2} i\mathbf{q} \cdot \left( \xi^{s'\dagger} \mathbf{S}_\chi \xi^s \right)\delta^{r' r} \,.
\end{align}

\bibliographystyle{JHEP}
\bibliography{ref,ref2}

\end{document}